\documentclass[]{article}
\usepackage[T1]{fontenc}
\usepackage[utf8]{inputenc}
\usepackage{lmodern}
\usepackage{microtype}
\usepackage{amsmath,amssymb,amsfonts}
\usepackage{mathtools} % Useful for cases* environment
\usepackage{bm}                        % bold math symbols
\usepackage{graphicx}
\usepackage{color}
\usepackage[usenames,dvipsnames]{xcolor}
\usepackage{booktabs}
\usepackage{multirow}
\usepackage{array}
\usepackage{tabularx}
\usepackage{float}
\usepackage{caption}
\usepackage{subcaption}
\usepackage{siunitx}                   % SI units
\usepackage{hyperref}
\usepackage{cleveref}
\usepackage{natbib}
\usepackage{geometry}
\usepackage{setspace}
\usepackage{abstract}
\usepackage{titlesec}
\usepackage{enumitem}

\title{%
  \textbf{Chirality-Driven Hierarchical Phase Morphologies in Self-Assembled Biaxial Amphiphiles}
}

\author{Sayantan Mondal and Jayashree Saha$^*$\\
\small Department of Physics, University of Calcutta, 
Kolkata, India\\
\small $^*$Correspondence: \texttt{jsphy@caluniv.ac.in}}

\date{\today}

\titleformat{\section}{\normalfont\bfseries\large}{\thesection.}{0.5em}{}
\titleformat{\subsection}{\normalfont\bfseries}{\thesubsection.}{0.5em}{}
\titleformat{\subsubsection}{\normalfont\itshape}{\thesubsubsection.}{0.5em}{}

\begin{document}
\maketitle

%========================================================
% Abstract
%========================================================
\begin{abstract}
\noindent
Chirality plays a crucial role in determining the structure of many systems in nature. Twisted or helical aggregates as a consequence of self-assembly can be seen in many biological and synthetic materials. Despite extensive theoretical and experimental efforts, how molecular-scale chirality gives rise to complex twisted morphologies in amphiphiles still remains unexplored. Here we study the interplay between molecular hydrophobicity, shape anisotropy and chirality using molecular dynamics simulation. Variation of relative molecular concentration and intrinsic chirality of molecules drive a sequence of twisted liquid crystalline variants of lamellar, cylindrical and vesicular phases. These structures emerge spontaneously under equilibrium conditions and are characterized by orientational correlation functions. We demonstrate that variation in molecular chirality gives rise to the development of hierarchical chiral order within the system. Further increment of chirality competes with hydrophobic interactions, leading to morphological instabilities. Our findings establish a direct link between microscopic chirality and mesoscale structure formation and their instabilities. Qualitative comparison of liquidity and pitch of the observed phase morphologies with the amount of chirality has been reported. 
\end{abstract}


\vspace{4pt}
\hrule
\vspace{8pt}

%========================================================
\section*{Introduction}
%========================================================
Self-assembly is the spontaneous organization of molecules into well-ordered structures. Amphiphilic molecules are capable of forming various spontaneous aggregates in aqueous solution. An amphiphile has a hydrophilic (water-attracting) head part and a hydrophobic (water-repelling) tail part. Due to this dual nature, amphiphiles can interact with polar as well as with non-polar substances. In the presence of water, the concentration of amphiphiles governs self-assembly into various lyotropic liquid-crystalline phases, including micelles, vesicles, cylinders, and flat bilayers. Single chain amphiphilic molecules(like detergents) are considered cylindrically symmetric, while molecules of biological importance such as cholesterols and phospholipids are highly asymmetric or non-cylindrical in nature. Biaxial ordering has been reported in amphiphilic systems\cite{Saupe,Saupe2,Akpinar,Reis,Neto}. Amphiphiles which are biaxial in nature, break the rotational degeneracy about the long axis, introducing additional orientational degrees of freedom. The coupling between biaxiality and chirality in the liquid crystalline systems has been the subject of extensive study{\cite{Lubensky,Kroin,Harris,Brand,Mondal}}.%These types of mesophases show different translational and orientational order with the temperature variation and they are fluidic in nature, Chirality is an asymmetry or broken improper rotational symmetry, which plays a significant role in diverse areas of nature. Molecules with inherent chirality can lead to the formation of helical superstructures in a typical liquid crystalline systems.
\\Mirror symmetry breaking governs the hierarchical organization of soft matter from lipid membranes to synthetic amphiphiles\cite{Castro,Tschierske,Fellows,Huang,Wang,Jiang,Alaasar}. In addition to their amphiphilic properties, chiral amphiphiles can exhibit broken improper rotational symmetry. Twisted morphologies driven by the self-assembly of chiral amphiphiles are ubiquitous in nature. Water-based self-assembly of chiral biomolecules led to the formation of nanohelices, ribbons, etc. via chirality transfer in supramolecular systems\cite{Caimi,Ouyang}. In lyotropic systems, chiral amphiphiles can generate twisted ribbons, helicoidal membranes and vesicular structures \cite{Schnur,Jonathan,Selinger,Sawa}. Propagation of structural asymmetry from the microscopic to macroscopic scale involves chirality transfer{\cite{Shaoxuan,Rong-Ming}} across multiple hierarchical levels. However, understanding the connection between molecular chirality and emergent mesoscale morphology remains a central challenge. Beyond chirality transfer in helical supramolecular systems, chiral amphiphilic self-assembly also plays a key role in tissue engineering, proteolytic stability, and cell apoptosis{\cite{Feng,Tang,Nanobiotechnol}}. Stereospecific permeability{\cite{Paegel}} and chiral recognition{\cite{Sato}} of the chiral lipid bilayer membrane is a major criterion for controlled release of drugs.
\\The self-assembly of chiral amphiphiles and their behavior have been extensively studied experimentally{\cite{Barclay,Ziserman,Zeng,Wei,Wu}}. Continuum theories{\cite{Selinger2,Seifert,Nyrkova}} establish the elastic framework and identify different stable morphologies theoretically. Yet, direct validation of these ideas through molecular-scale models remains limited. Computer simulation studies are essential in order to understand the mesoscopic phase behavior by chiral amphiphilic molecules via self-assembly. For a fully hydrated system, the presence of explicit solvent would take up a lot of computational time. This further motivates the need for an implicit-solvent coarse-grained model for mesoscale simulation of chiral amphiphiles in bulk water. Instead of considering the water molecules explicitly, solvent-free{\cite{Cooke,Deserno,Ugarte}} models take its presence into account by recognizing the hydrophobic effect as an attractive force between the amphiphiles. Hydrophobicity is accounted for by the broad-range attractive potential between tails. On the other hand, the hydrophilic beads act merely as hard spheres providing the excluded volume interaction.  
\\Our implicit-solvent model captures the hydrophobic driving force for molecular aggregation while incorporating molecular chirality through an additional chiral term in the pair potential. To the best of our knowledge, this is the first molecular framework that simultaneously incorporates molecular shape anisotropy, inherent chirality, and amphiphilic aggregation. This model produces chirality-induced bilayer morphologies including lamellar, cylindrical, and vesicular phases. In this work, we present a molecular dynamics(MD) mesoscale simulation of chiral amphiphiles using minimal coarse-graining by representing each amphiphile with two bead segments, spherical head bead and rod-like elongated tail bead(biaxial ellipsoid). Our study allows us to understand the role of inherent molecular chirality on the structure of generated phase morphologies by amphiphilic self-assembly, and also how the chiral self-assembly leads to the morphological instability and frustrated mesophases.   

%========================================================
\section*{Model and Methods}
%========================================================
For ellipsoidal geometry of the molecules, ellipsoid contact function(ECF) was used{\cite{Perram,Paramonov}}. Ellipsoid contact potential(ECP){\cite{Perram_et_al}} which is based on ECF, can well describe highly anisotropic systems. Here, we have used modified ECP{\cite{saha}} in implicit solvent{\cite{Bhowmick}} with chiral interaction{\cite{Mondal}} to study the self-assembly of identical biaxial chiral amphiphiles.
\subsection*{Coarse-grained model}
Each molecule is represented by two segments, one is a spherical head and another is an ellipsoidal tail bead, rigidly attached to one another. The solvent-mediated interaction between molecular tail beads is represented by a broad-range attractive potential{\cite{Cooke}}. The size of each molecular segment is defined by a repulsive excluded-volume potential.
\begin{equation}
V_{repulsive} = 
\begin{cases}
4\epsilon\left[ \rho^{-12} - \rho^{-6} + \frac{1}{4} \right], & r \leq r_c \\
0                                                           , & r > r_c
\end{cases}
\end{equation}  
Here, $r_c=2^{\frac{1}{6}}\sigma_0-\sigma_0+\sigma$ and $$\rho=\left(\frac{r_{ij}-\sigma+\sigma_0}{\sigma_0}\right)$$. The intermolecular distances between the centre of masses of two molecular ellipsoids is given by $r_{ij}$. The smallest semi-axes length of ellipsoidal tail bead and the diameter of spherical head bead is $\sigma_0$, which is the length scale of our model potential. $\epsilon$ is the total well-depth function. The orientation-dependent energy strength \& distance functions are given by, $\epsilon\equiv\epsilon(\hat{a}_{i},\hat{a}_{j},\hat{r})$ and $\sigma\equiv\sigma(\hat{a}_{i},\hat{a}_{j},\hat{r})$. $\hat{a}_{i}$ and $\hat{a}_{j}$ are rotational matrices related to the molecular orientations in 3-d euclidean space and used for the transformation of molecules $i$ and $j$ from laboratory frame to body frame coordinates.
\\The absence of explicit solvent molecules and their hydrophobicity are accounted for by an attraction between tail-beads. The potential with a flat width $w_f$ at their minima represented by,
\begin{equation}
V_{attract} = 
\begin{cases}
-\epsilon                                               , & r < r_c+w_f \\
4\epsilon\left[ \rho^{-12}_{_f} - \rho^{-6}_{_f} \right], & r_c \leq r \leq w_f + w_c \\
0                                                       , & r > w_f + w_c
\end{cases}
\end{equation}
Where, $${\rho}_f=\left(\frac{r_{ij}-\sigma+\sigma_0-w_f}{\sigma_0}\right)$$. This solvent mediated interaction is cut off beyond $w_f+w_c$.
\\For molecular chirality, we have used a pseudo-scalar term to represent the chiral interaction potential{\cite{van der meer,paul}} working between tail-beads.
\begin{equation}
V_{chiral} = 4c\epsilon\rho^{-7} [(\hat{a}_{i3}\times\hat{a}_{j3})\cdot{\hat{r}_{ij}}](\hat{a}_{i3}\cdot\hat{a}_{j3})
\end{equation}
Here, $\hat{a}_{i3}$ and $\hat{a}_{j3}$ are the unit vectors along molecular semi-major axis of $i$'th and $j$'th molecules respectively. This term produces the intermolecular torque between neighbouring molecules and is responsible for the handedness of the system. $c$ is the chiral strength parameter. For achiral molecules, $c=0$ and for a system of chiral molecules the amount of $c$ is non-zero.
\\The total interacting pair potential between neighbouring molecules is given by,
\begin{equation}
V_{total} = V_{repulsive} + V_{attract} + V_{chiral}
\end{equation}
The anisotropic pair potential above accounts for biaxial contributions in both shape and energy of the ellipsoidal tail beads.\\The analytical expressions for the orientation-dependent energy strength and distance functions of the ECP, together with the derived forces and torques, are provided in Appendix~\ref{app:ecp}. 
%============================================================================
\subsection*{Simulation details}
Simulations were performed in the canonical(NVT) ensemble using Nose-Hoover thermostat{\cite{holian}} with periodic boundary conditions. For the integration of the equation of motions{\cite{Allen}}, leap-frog verlet scheme was implemented. To check for the system size effect, different system sizes of $N=864, 1372, 2048$ have been considered. The results obtained were qualitatively similar. In this article, we have presented the results of a system containing $N=1372$ molecules. Each molecule consists of two beads, where the head bead is spherical in shape and the tail bead is a biaxial ellipsoid. The diameter of the spherical head bead and the smallest semi-axis length of ellipsoidal tail has been chosen to be $\sigma_0$, which is the length scale of our model potential. The size ratio $ \sigma_x \colon \sigma_y \colon \sigma_z $ of the biaxial ellipsoidal tail were considered as $ 1 \colon 1.5 \colon 3.0 $, where $ \sigma_x = \frac{\sigma_0}{2} $. The scaled moment of inertia($I^{*} = \frac{I}{m{\sigma_0}^2}$) values were correspondingly $I^*_{xx} = 0.5625,I^*_{yy} = 0.50, I^*_{zz} = 0.1625$. The well-depth energy ratios of the biaxial ellipsoidal tail were taken as $ \epsilon_x \colon \epsilon_y \colon \epsilon_z = 1 \colon (1.0/2.25) \colon (1/5.80) $, where $ \epsilon_x = \epsilon_0 = 1 $. For the spherical head bead of each molecule, $ \sigma_x \colon \sigma_y \colon \sigma_z = 1 \colon 1 \colon 1$ and $ \epsilon_x \colon \epsilon_y \colon \epsilon_z = 1 \colon 1 \colon 1 $. The well-depth parameters were $\mu=2$ \& $\nu=1$ and mass $m = 1$. 
\\ Initially, molecules were placed on an FCC lattice within a cubic simulation box with periodic boundary conditions and minimum image convention, with randomized orientations and positions. Initial translational and angular velocities were drawn from a Gaussian(Maxwell-Boltzmann) distribution. Proper tuning of the effective molecular concentration was achieved by systematically varying the simulation box size. We have simulated our system for different lengths of the box sizes ranging from $L_{box} = 30\sigma_0$ to $43\sigma_0$. By changing the length of the box, different phase morphologies have been observed by the self-assembly of biaxial amphiphilic molecules. Flat bilayer phases when $L_{box} = 31\sigma_0$ to $34\sigma_0$, in between $L_{box} = 35\sigma_0$ to $39\sigma_0$ cylindrical bilayer phases and for $L_{box}>40\sigma_0$ vesicle bilayer phases have been observed. To generate an isotropic initial configuration, the system was melted at a high reduced temperature of $T^{*}(\equiv {K_{B}T}/{\epsilon_0})=3.0$ for each box size. After that, this equilibrated rotational and positional isotropic molecular configuration served as the initial configuration of our simulation studies in order to observe self-assembled amphiphilic phase behavior. Each simulation run was initialized from a configuration equilibrated at a higher temperature, after which $T^{*}$ was gradually decreased to explore the ordered phase structures. The cut-off radius was chosen to be $R_{cut}=5.0\sigma_{0}$, beyond this cut-off distance our anisotropic model potential was truncated. By changing $R_{cut}$, the phase stability occurs at a slightly different temperature. However, the results remained qualitatively similar. During cooling, a time step of $dt = 0.001$ was used near the ordered state. The long-range orientational order of the mesophases is adequately captured using this cutoff distance in conjunction with the potential's shifted form. To represent the orientation of each molecule in 3-d Euclidean space, we used quaternions.
%For the characterisation of the self-assembled phases formed by biaxial amphiphiles
%simulation run started from complete isotropic configuration
\subsection*{Results and Analysis}
\begin{figure*}[htbp]
    \centering
    \begin{subfigure}[c]{3cm}
        \includegraphics[width=\textwidth]{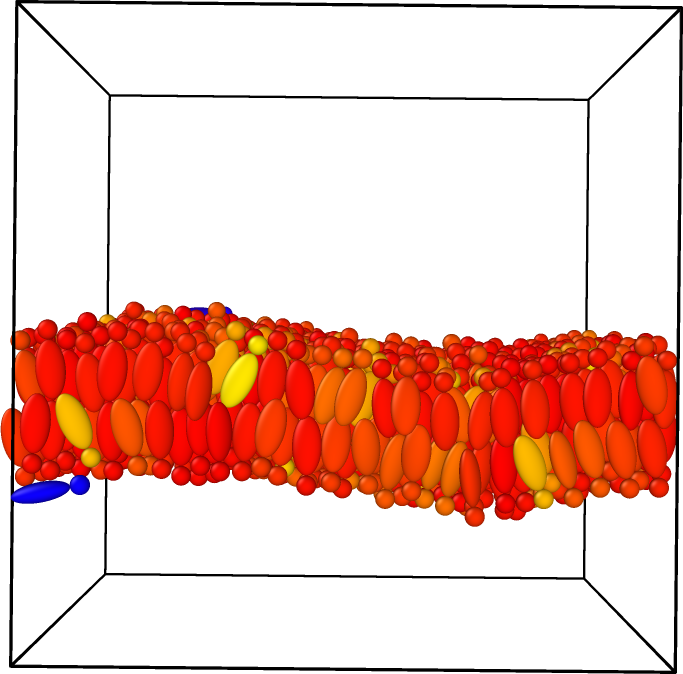}
        \caption{$c=1.50$, side view}
    \end{subfigure}
    %\hfill
    \begin{subfigure}[c]{3cm}
        \includegraphics[width=\textwidth]{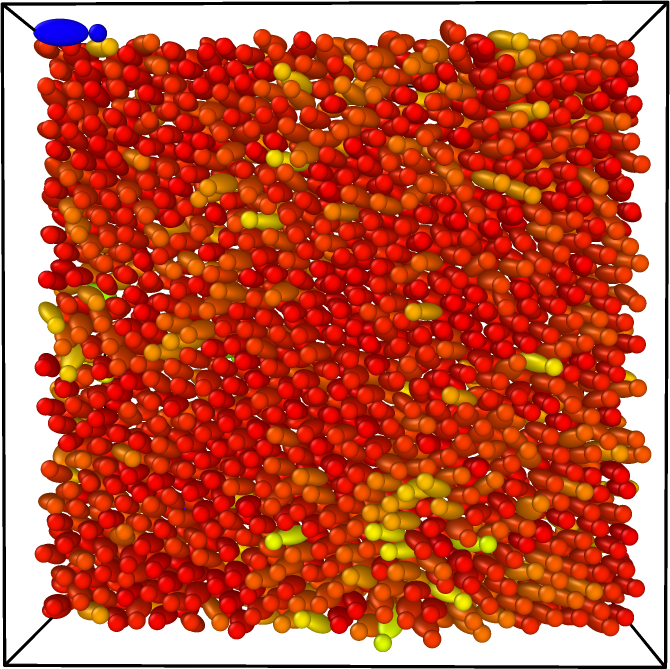}
        \caption{$c=1.50$, top view}
    \end{subfigure}
    %\hfill
    \begin{subfigure}[c]{3cm}
        \includegraphics[width=\textwidth]{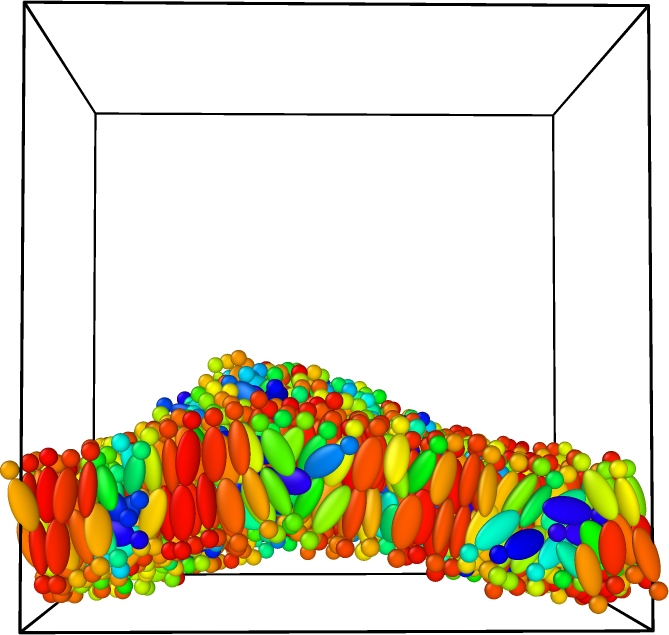}
        \caption{$c=2.0$, side view}
    \end{subfigure}
    %\hfill
    \begin{subfigure}[c]{3cm}
        \includegraphics[width=\textwidth]{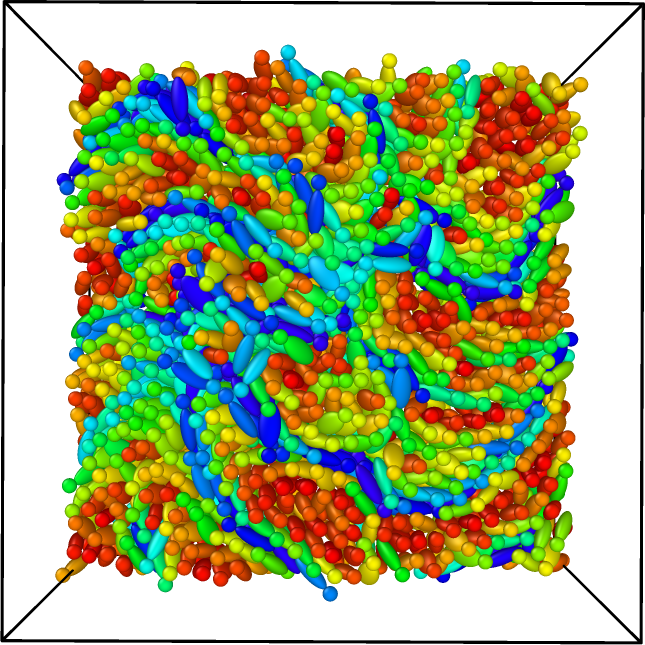}
        \caption{$c=2.0$, top view}
    \end{subfigure}
    %\hfill
    \begin{subfigure}[c]{3cm}
        \includegraphics[width=\textwidth]{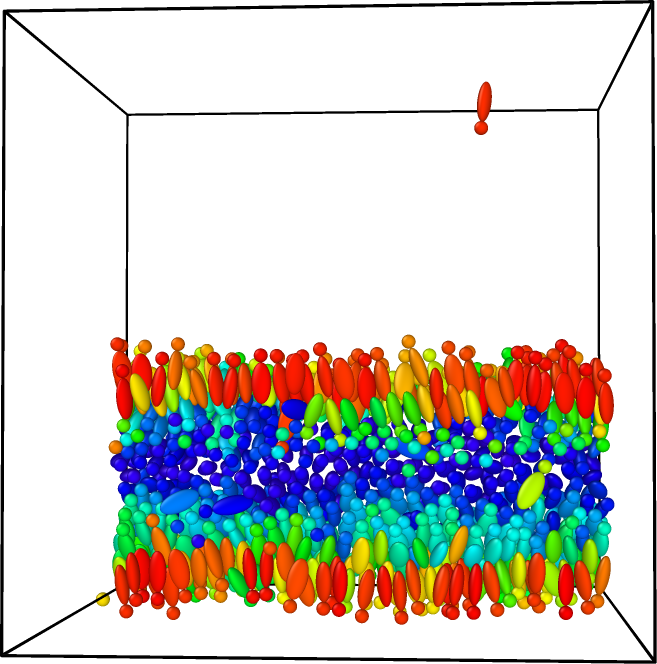}
        \caption{$c=0.0$, cut-view}
    \end{subfigure}
    \begin{subfigure}[c]{3cm}
        \includegraphics[width=\textwidth]{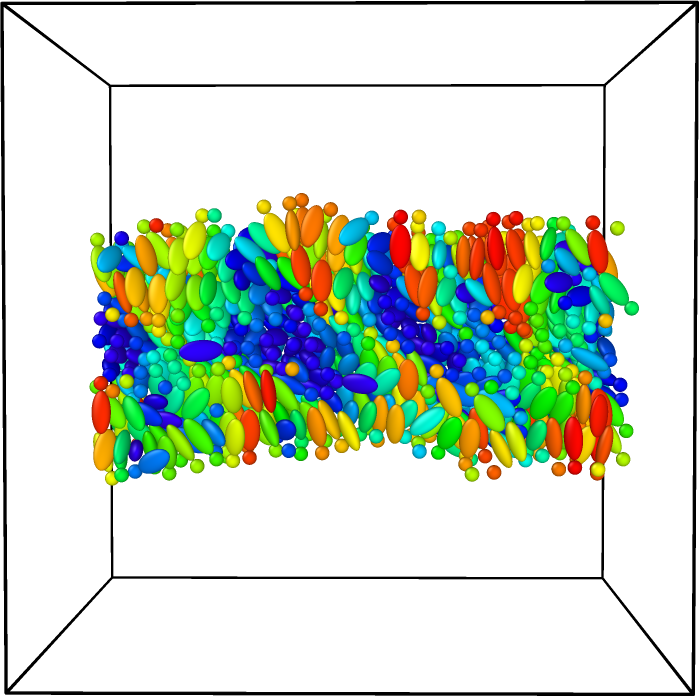}
        \caption{$c=2.0$, cut-view inner surface}
    \end{subfigure}
    \begin{subfigure}[c]{3cm}
        \includegraphics[width=\textwidth]{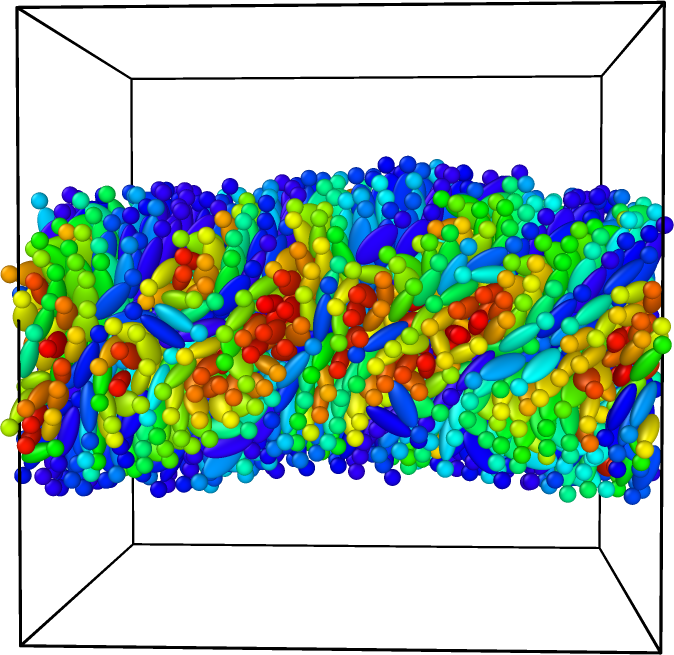}
        \caption{$c=2.0$, outer surface}
    \end{subfigure}
    \begin{subfigure}[c]{3cm}
        \includegraphics[width=\textwidth]{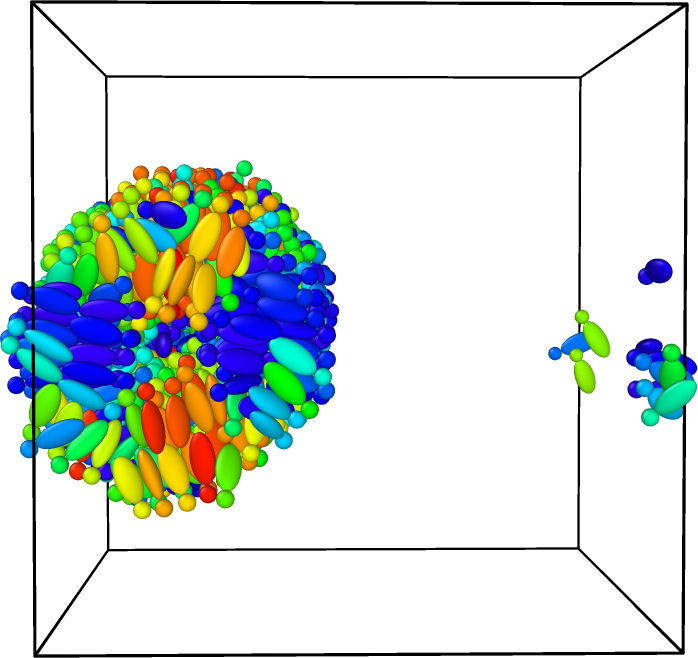}
        \caption{$c=2.0$, top view}
    \end{subfigure}
    \begin{subfigure}[c]{3cm}
        \includegraphics[width=\textwidth]{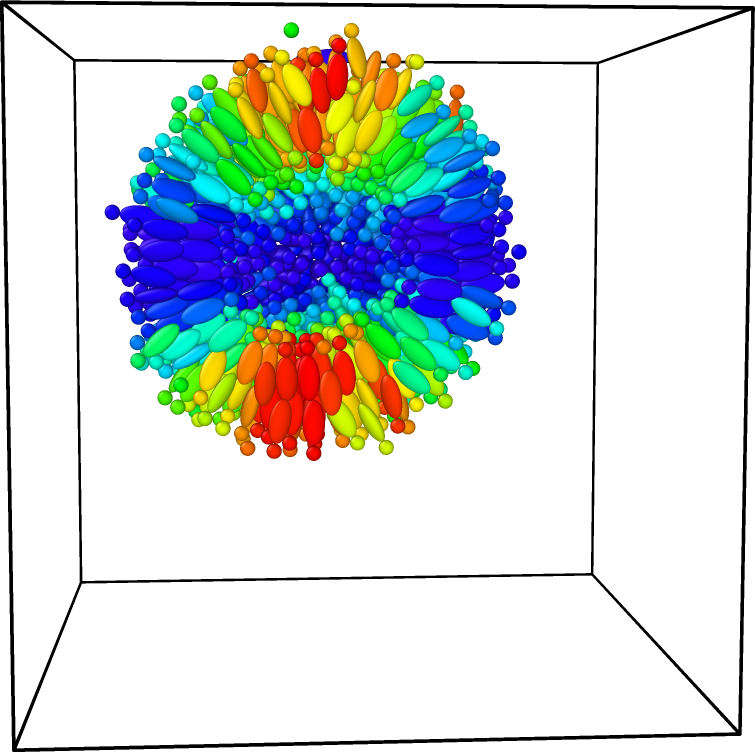}
        \caption{$c=0.0$, cut-view}
    \end{subfigure}
    \begin{subfigure}[c]{3cm}
        \includegraphics[width=\textwidth]{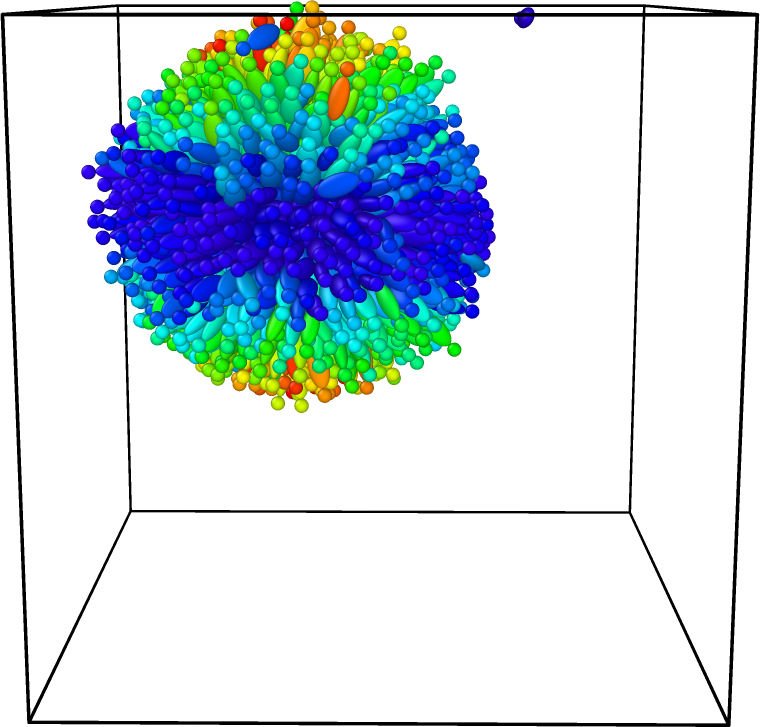}
        \caption{$c=0.0$}
    \end{subfigure}
    \begin{subfigure}[c]{3cm}
        \includegraphics[width=\textwidth]{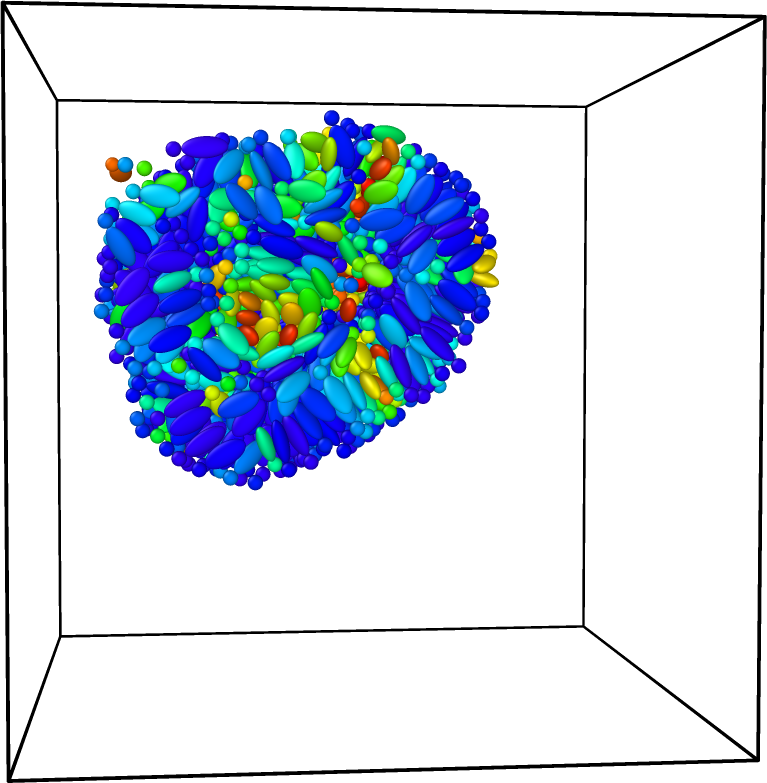}
        \caption{$c=2.0$, cut-view}
    \end{subfigure}
    \begin{subfigure}[c]{3cm}
        \includegraphics[width=\textwidth]{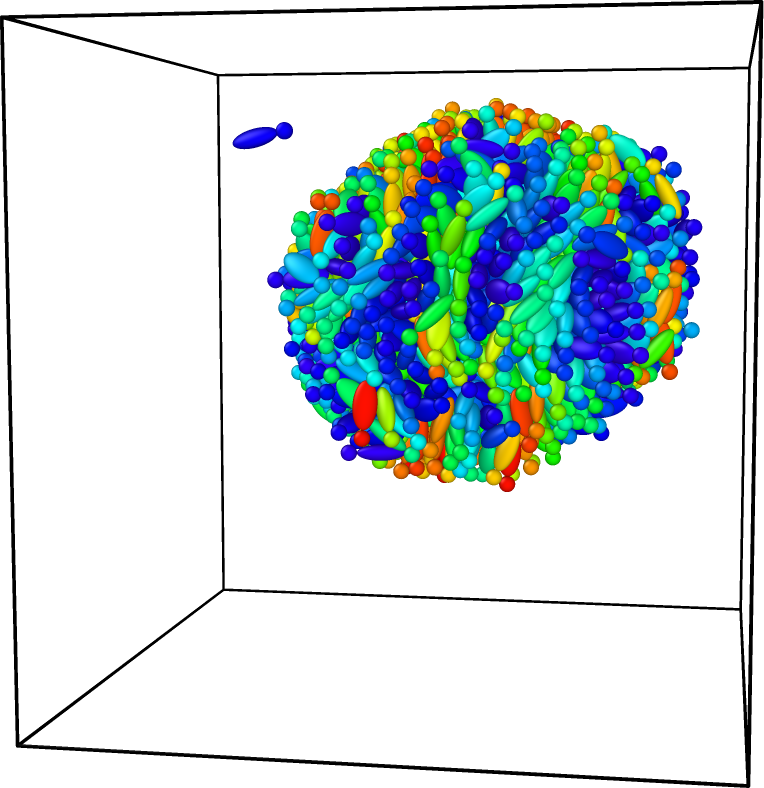}
        \caption{$c=2.0$}
    \end{subfigure}
    \caption{Snapshots of molecular configurations for (a–d) lamellar, (e–h) cylindrical, and (i–l) vesicular bilayer phases at selected values of chiral strength $c$.}
\end{figure*}
To study chirality-driven self-assembly in biaxial amphiphiles, we performed implicit-solvent molecular dynamics simulations of a system of N = 1372 biaxial chiral amphiphiles. The value of the chiral strength parameter($c$) was varied for different sizes of the box length($L_{box}$). With gradual increase in molecular chirality, director field twisting is observed in flat two-dimensional bilayer stacks and in curved surfaces like cylinder and vesicle phases. Snapshots of the molecular configurations of the phases observed are presented in Fig.$1$. For visualization, the squared components of the molecular semi-major axis($\hat{a}^{2}_{3x}$, $\hat{a}^{2}_{3y}$, $\hat{a}^{2}_{3z}$) were mapped to RGB color channels to represent directional information in OVITO, color-coded according to the normal unit vectors for each configuration. Representative snapshots at selected values of $c$ are presented to illustrate the phases formed by the chiral amphiphilic system. In the absence of chirality, an ordinary flat lamellar phase was observed at $L_{box}=33$. When $c=1.5$, wavy bilayer phase was observed(Fig.$1a$). Further increase in chirality amplifies the waviness, eventually disrupting the bilayer structure. A modulated helical network(Fig.$1d$) has been observed with helical director field modulation. For $L_{box}=35.5$, cylindrical bilayer(Fig.$1e$) has been observed. Gradual increase in chiral strength($c$) leads to the creation of a twisted cylindrical phase, where a helical field-modulated twist forms along the cylindrical axis. The inner cross-sectional view of the cylinder along axial direction in Fig.$1f$ and outer surface in Fig.$1g$ confirmed small chiral domains along cylinder axis keeping their bilayer structure intact. At $L_{box}=42$, bilayered vesicle(Fig.$1i$ \& Fig.$1j$) was generated. As $c$ increases, the vesicle starts to deviate from its spherical shape. The helical modification of the bilayered structure takes place in vesicle. All phases formed at higher chiral strengths($c \ge 2.0$) are frustrated mesophases. To assess the role of molecular biaxiality, we compared the present results with those obtained for the corresponding uniaxial(tail) amphiphilic model under similar thermodynamic conditions. While the uniaxial system($ \sigma_x \colon \sigma_y \colon \sigma_z  \equiv  1 \colon 1 \colon 3 $ ; $ \epsilon_x \colon \epsilon_y \colon \epsilon_z = 1 \colon 1 \colon (1/5) $) exhibited conventional self-assembled structures, we did not observe twisted lamellae, cylindrical bilayers, or vesicular bilayers within the parameter range explored. This comparison suggests that molecular biaxiality is a crucial ingredient for coupling chirality to collective membrane deformations and stabilizing the complex morphologies observed in the present study.
\\These frustrated structures arise from competing orientational preferences between hydrophobic interactions and intrinsic molecular chirality. Further increasing $c$ brings more instability as a consequence of the competing interactions, which leads to the complete disruption of bilayer structure.
\\ After reaching equilibrium in an ordered state, we were able to examine the density profiles of head \& tail beads with radial distance. In case of flat bilayer this radial distance was measured along the layer normal, while for vesicle radial distance is measured from its center and for cylinder, radial distance is perpendicular to its own axis. 
\begin{figure*}[htbp]
    \centering
    \begin{subfigure}[b]{5cm}
        \includegraphics[width=\textwidth]{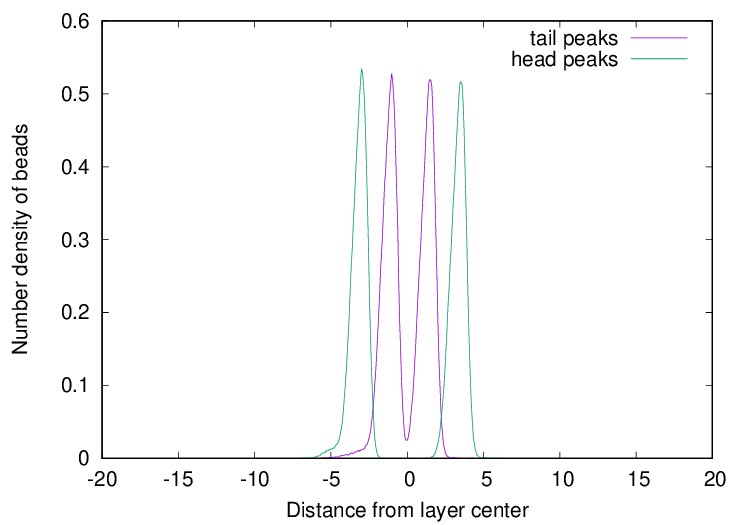}
        \caption{$c=1.0$, lamellar}
    \end{subfigure}
    %\hfill
    \begin{subfigure}[b]{5cm}
        \includegraphics[width=\textwidth]{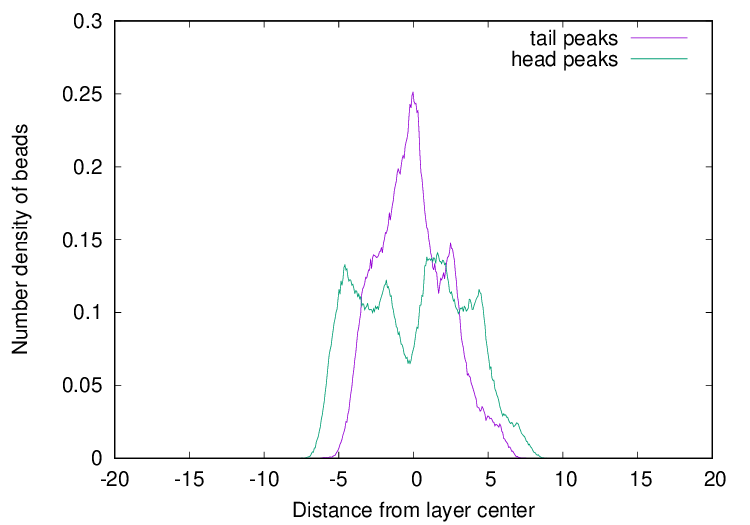}
        \caption{$c=2.0$, lamellar}
    \end{subfigure}
    %\hfill
    \begin{subfigure}[b]{5cm}
        \includegraphics[width=\textwidth]{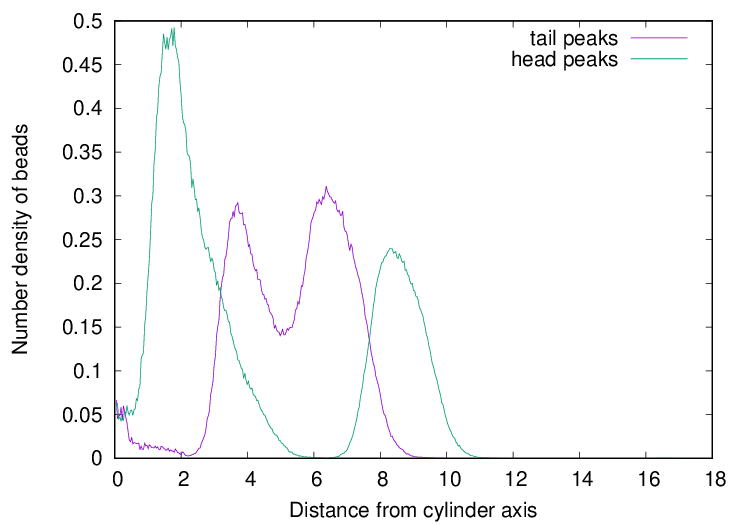}
        \caption{$c=1.0$, cylindrical}
    \end{subfigure}
    %\hfill
    \begin{subfigure}[b]{5cm}
        \includegraphics[width=\textwidth]{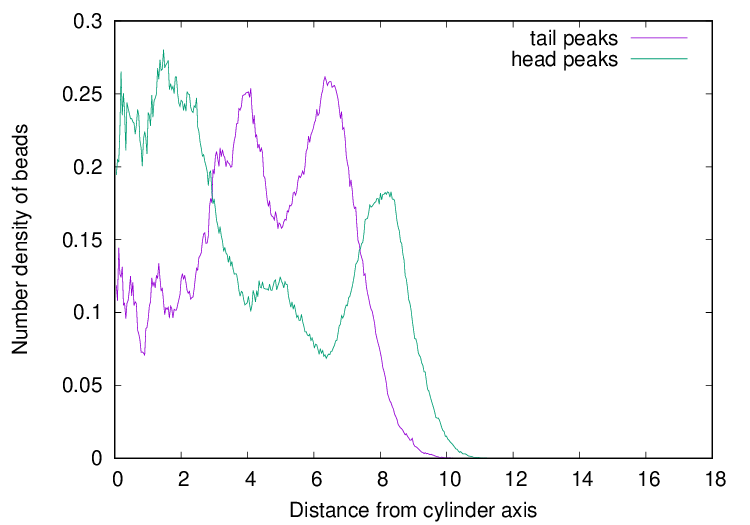}
        \caption{$c=2.50$, cylindrical}
    \end{subfigure}
    %\hfill
    \begin{subfigure}[b]{5cm}
        \includegraphics[width=\textwidth]{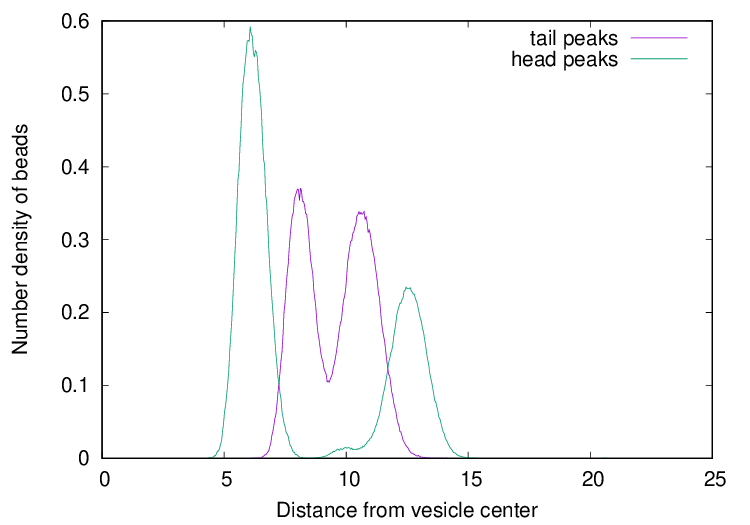}
        \caption{$c=1.0$, vesicular}
    \end{subfigure}
    %\hfill
    \begin{subfigure}[b]{5cm}
        \includegraphics[width=\textwidth]{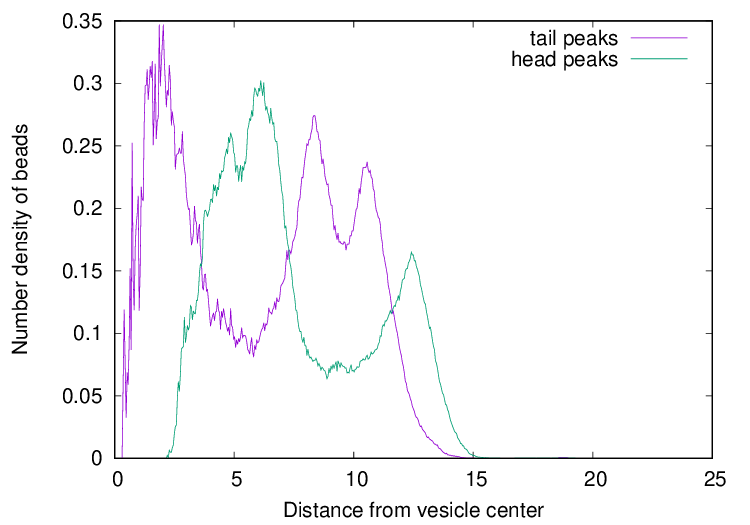}
        \caption{$c=2.0$, vesicular}
    \end{subfigure}
    \caption{Density peaks at radial distances, (a–b) are lamellar, (c–d) cylindrical, (e–f) vesicular.}
\end{figure*}
A typical plot of such a density proﬁle is shown in Fig.$2$, where the number densities of head and tail beads in both leaflets are clearly visible. Increasing chirality destabilizes the geometric structure of each morphology, which is clearly evident by the peaks smearing out at higher chirality. For flat lamellar and vesicle phase at $c \geq 2.0$, this layered structure is disrupted. However, the cylindrical bilayer retains its layered structure for $c < 2.50$, beyond which a secondary twist develops. This happens due to a competitive interaction between the hydrophobicity and inherent molecular homochirality of an amphiphilic molecule. When the chiral strength parameter is low then hydrophobicity of an amphiphile dominates, creating achiral phases where no net chirality formed in liquid phases. At higher values of chirality, intermolecular twisting capability dominates over their hydrophobic interactions leading to frustrated chiral mesophases. Lamellar phase becomes wavy at lower chirality, and the waviness increases with increasing chirality up until the bilayer structures get completely disrupted. However, this chirality-driven instability manifests differently across different morphologies. The reason lies in their geometrical arrangement. In a flat bilayer, molecules are arranged with their tails in an antiparallel configuration. When molecular chirality is introduced, each molecule exerts a torque on its neighbors. The neighboring molecule cannot complete a full rotation while maintaining its hydrophobic tail-tail core intact, resulting in a wavy two-dimensional bilayer sheet. While in a cylindrical bilayer, along the cylinder axis molecules can complete a rotation with respect to its neighbours keeping their bilayer tail-tail core intact without disrupting till a certain threshold value is reached, creating a helical network with small chiral domains on the outer surface of the cylinder. For the vesicle, molecules have the freedom to rotate in $3d$ radially outward from the center, the cost of forming a three-dimensional helical network is the disruption of the hydrophobic tail-tail core, which destabilizes the bilayer; although the macroscopic helicity in vesicle bilayer system is not uniform in all directions.
\\ Orientational correlations were investigated with the help of a scalar longitudinal orientational correlation function{\cite{Stone}} $S_{220}({r^{*}_{||}})$ and a pseudo-scalar longitudinal orientational correlation function $S_{221}({r^{*}_{||}})$. These correlation functions were measured as a function of intermolecular distance ${r^{*}_{||}}$ along a reference axis. The mathematical expressions of the correlation functions are given by,
\begin{equation}
S_{220}({r^{*}_{||}}) = {\frac{1}{2\sqrt{5}}}{\langle 3(\hat{a}_{i3}\cdot\hat{a}_{j3})^{2}-1 \rangle}
\end{equation}
\begin{equation}
S_{221}({r^{*}_{||}}) = -\sqrt{\frac{3}{10}}{\langle \{(\hat{a}_{i3}\times\hat{a}_{j3})\cdot{\hat{r}_{ij}}\}(\hat{a}_{i3}\cdot\hat{a}_{j3}) \rangle}
\end{equation}
\begin{figure*}[htbp]
    \centering
    \begin{subfigure}[b]{4cm}
        \includegraphics[width=\textwidth]{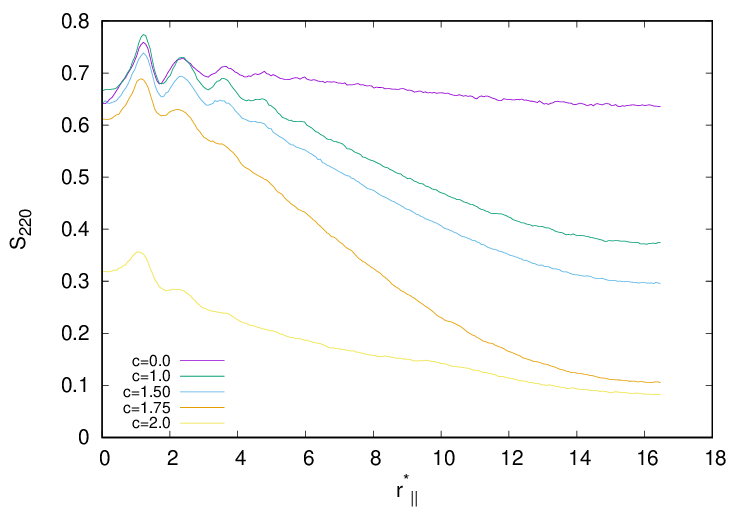}
        \caption{$S_{220}$ in layer plane}
    \end{subfigure}
    %\hfill
    \begin{subfigure}[b]{4cm}
        \includegraphics[width=\textwidth]{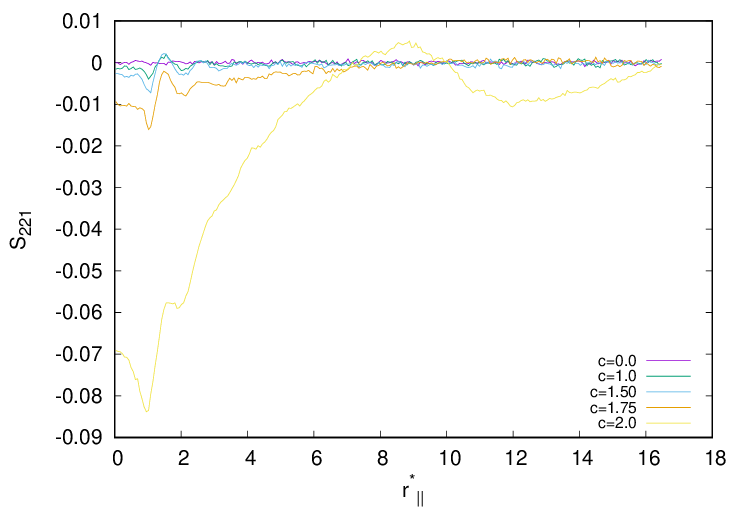}
        \caption{$S_{221}$ in layer plane}
    \end{subfigure}
    %\hfill
    \begin{subfigure}[b]{4cm}
        \includegraphics[width=\textwidth]{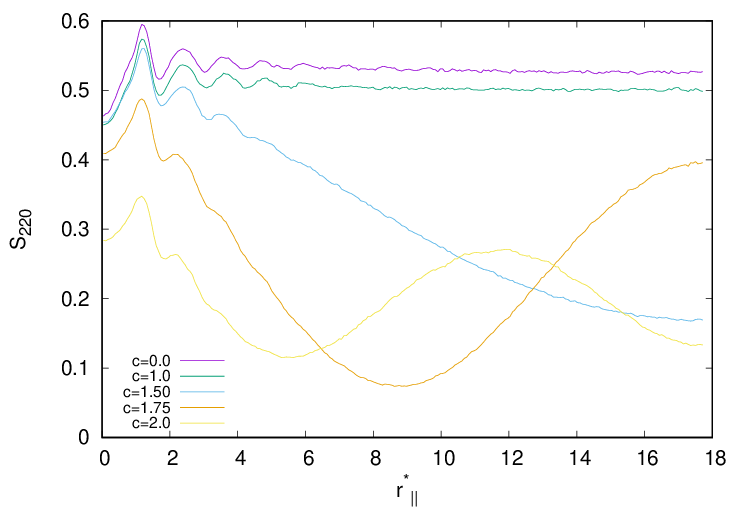}
        \caption{$S_{220}$ along cylindrical axis}
    \end{subfigure}
    %\hfill
    \begin{subfigure}[b]{4cm}
        \includegraphics[width=\textwidth]{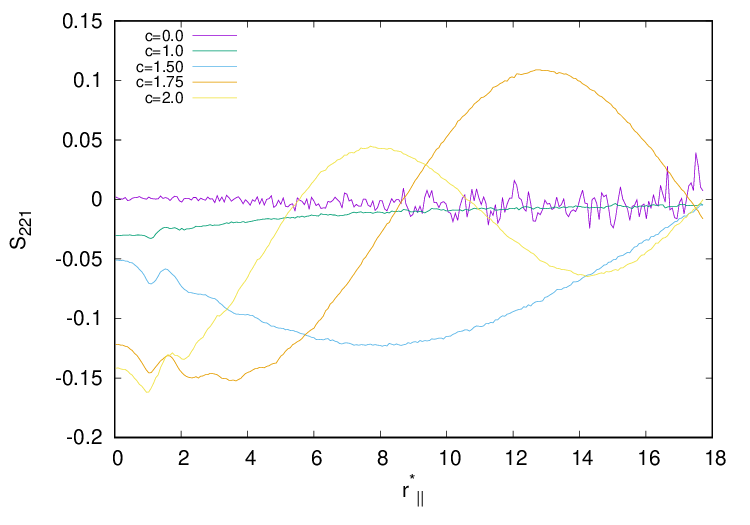}
        \caption{$S_{221}$ along cylindrical axis}
    \end{subfigure}
    %\hfill
    \begin{subfigure}[b]{4cm}
        \includegraphics[width=\textwidth]{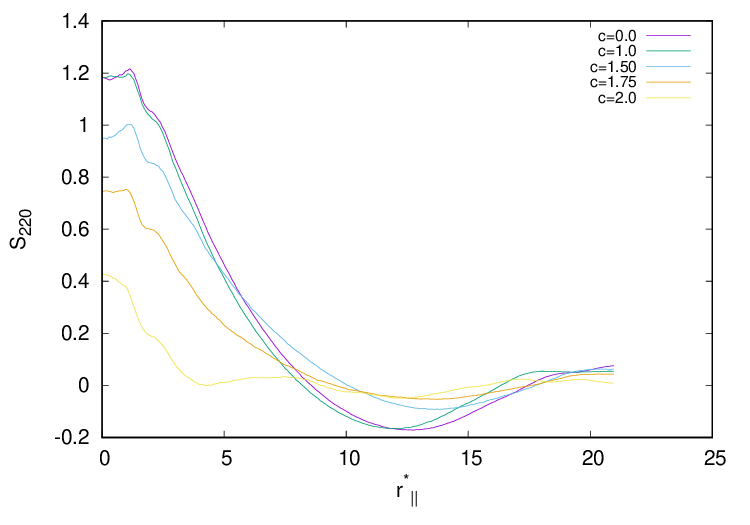}
        \caption{$S_{220}$ along radial axis}
    \end{subfigure}
    %\hfill
    \begin{subfigure}[b]{4cm}
        \includegraphics[width=\textwidth]{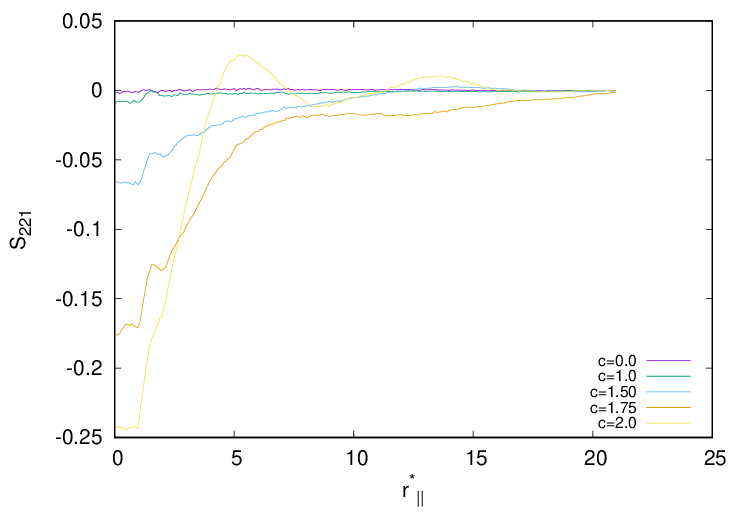}
        \caption{$S_{221}$ along radial axis}
    \end{subfigure}
    \caption{Orientational correlation functions for (a), (b) lamellar, (c),(d) cylinder, (e), (f) vesicle.}
\end{figure*}
\begin{figure*}[htbp]
    \centering
    \begin{subfigure}[b]{4cm}
        \includegraphics[width=\textwidth]{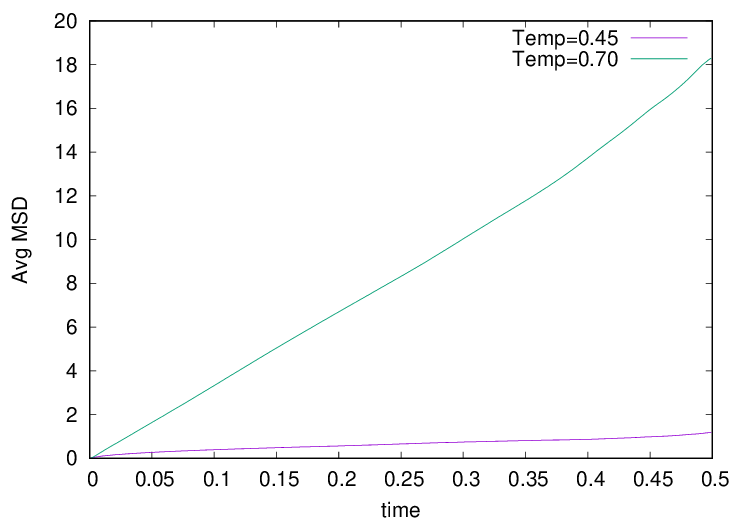}
        \caption{Twisted lamellar, $c=2.0$}
    \end{subfigure}
    %\hfill
    \begin{subfigure}[b]{4cm}
        \includegraphics[width=\textwidth]{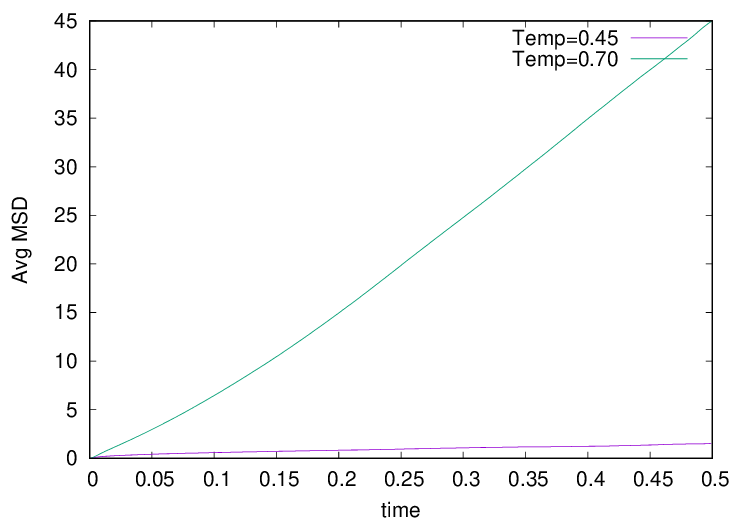}
        \caption{Twisted cylinder, $c=2.0$}
    \end{subfigure}
    %\hfill
    \begin{subfigure}[b]{4cm}
        \includegraphics[width=\textwidth]{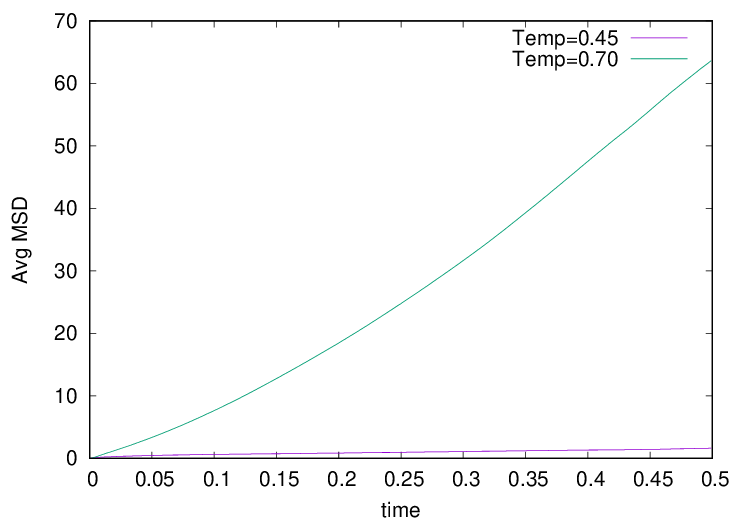}
        \caption{Twisted vesicle, $c=2.0$}
    \end{subfigure}
    %\hfill
    \begin{subfigure}[b]{4cm}
        \includegraphics[width=\textwidth]{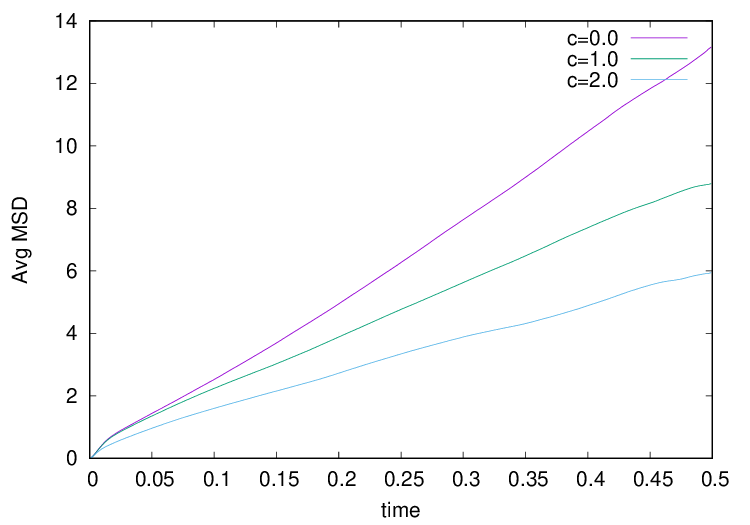}
        \caption{Twisted cylinder at varying $c$}
    \end{subfigure}
    \caption{MSD variations with time for twisted amphiphilic phases}
\end{figure*}
\begin{table*}[t]
\centering
\caption{Diffusion coefficient $D_{\text{VACF}}$ at reduced temperature of $T^{*}=0.70$}
\begin{tabular}{c c c c c}
\toprule
Morphology & $c$ & $D_{\text{VACF}}$ \\
\midrule
\textbf{Lamellar} & 0  & 3.12E-003 \\
\allowbreak & 1  & 2.26E-003 \\
\allowbreak & 2  & 9.69E-004 \\
\textbf{Cylinder} & 0  & 1.52E-003 \\
\allowbreak & 1  & 1.47E-003 \\
\allowbreak & 2  & 1.11E-003 \\
\textbf{Vesicle} & 0  & 8.22E-003 \\
\allowbreak & 1  & 5.89E-003 \\
\allowbreak & 2  & 1.16E-003 \\
\bottomrule
\end{tabular}
\end{table*}
The scalar correlation function $S_{220}$ can measure the tendency of a molecule to align parallel with its neighbour, it attains maximum when the long axes(semi-major) of the neighbouring molecules are parallel. While the pseudo-scalar correlation function $S_{221}$ becomes non-zero only in systems which have broken improper rotational symmetry, serving as a direct evidence of chirality. This pseudo-scalar term, which is a consequence of the triple product in its definition, can change sign under mirror reflection. To evaluate ensemble average of the above mentioned correlation functions, after reaching equilibrium we have taken $500$ different molecular configurations at each $100$ md steps. For a pair of molecules separated by a distance $r^{*}_{||}$ along a chosen reference axis considering minimum image convention the function $S_{220}$ reached its maximum value when the semi-major axes of two neighbouring amphiphilic molecules were parallel to one another, whereas $S_{221}$ attained an extremum value when molecular semi-major axes were at a $45^{\circ}$ angle to each other. In order to investigate chirality driven correlations on a system of biaxial amphiphiles, we calculated the above functions for different morphological phases like lamellar, cylindrical and vesicular bilayer for different values of chiral strength parameter $c$. These orientational correlation functions were measured along three mutually perpendicular axis of the simulation box and along different director axes. The plots in Fig.$3$ reveal the orientational order of different morphologies by varying chiral strength. In lamellar bilayer structure, in plane parallel to the $2-d$ bilayer sheet the function $S_{220}$(in fig.$3a$) for achiral case($c = 0$) shows, a flat correlation indicating strong parallel orientational order of flat bilayer phase. Whereas for achiral system $S_{221}$(in fig.$3b$) remains zero showing expected behavior. As chirality increases($c > 1.0$), $S_{221}$ develops very small non-zero oscillatory behavior, without developing a significant helical ordering. Further increasing chiral strength $c$ leads to more pronounced sinusoidal variations in only $S_{221}$ for lamellar phase. However, since a $2-d$ stack of bilayer sheet tries to retain its layer structure due to the hydrophobicity, $S_{220}$ does not reach a second peak which suggests anti-parallel configuration. So further increasing chirality leads to the disruption of bilayer structure. In cylindrical bilayer structure, along the axis of the cylinder $S_{220}$ \& $S_{221}$(in fig.$3c-3d$) shows similar behavior as that of lamellar phase when $c<1.50$, and sinusoidal oscillations exhibited at higher chirality. With increasing chirality along the axial direction of the cylinder, the spatial period of oscillation decreases and more peaks trying to fit within the length of the box, reducing the pitch significantly. The oscillating nature in cylindrical morphology for both $S_{220}$ and $S_{221}$ functions is suggestive of the enhancement of chiral ordering. For vesicular bilayer phase, orientational correlation along mutually perpendicular axes(in fig.$3e-3f$) shows similar variations due to its closed symmetric nature. However at $c=2.0$, the pseudo-scalar function $S_{221}$ shows damped oscillating behavior, indicating the development of a weak helical ordering. 
\\Fluidity of the observed phase morphologies is demonstrated in Fig.$4$ with the help of mean square displacements(MSD) of the centers of mass(molecular tails). After equilibration from a typical run, obtained MSD($\langle\Delta r^2\rangle$) over time provides evidence on the translational dynamics of the observed phases. The variation of MSD with time, while decreasing the temperature from $T^{*}=0.70$ to $T^{*}=0.45$, liquidity became restricted due to strong orientational order at low temperature. To calculate diffusivity, the diffusion constant $D$, we utilized Green-Kubo's relation{\cite{Allen,Frenkel}} for $2d$ and $3d$ diffusion correspondingly in planar and curved morphologies. The values of diffusion coefficient with the change of chirality are reported in Table$1$, and with increasing value of chiral strength parameter the value of the diffusion coefficient decreases. Also, Fig.$4d$ shows for cylindrical phase, the slope of MSD variation with time for a particular temperature($T^{*}=0.55$) decreases by increasing molecular chirality. Similar behavior is obtained for lamellar as well as vesicular phases. So, the MSD graphs and calculated diffusivity implies, as the intrinsic chirality of a molecule grows the structural phase morphologies become less mobile.%along the plane parallel to radial direction the funtion $S_{220}$ shows a first peak for neighbouring parallel molecular configuration then it gradually decays down with distance for every values of $c$ showing a typical molecular alignment of a cylinder circular face geometry. Whereas, $S_{221}$ shows a very weak damped oscillatory behavior when $c>1.0$. A
\begin{figure*}[htbp]
    \centering
    \begin{subfigure}[b]{5cm}
        \includegraphics[width=\textwidth]{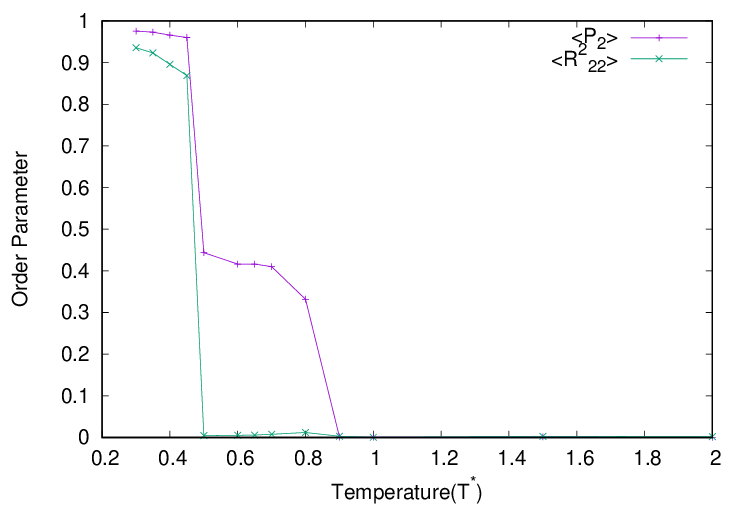}
        \caption{Flat bilayered phase, $c=0.0$}
    \end{subfigure}
    %\hfill
    \begin{subfigure}[b]{5cm}
        \includegraphics[width=\textwidth]{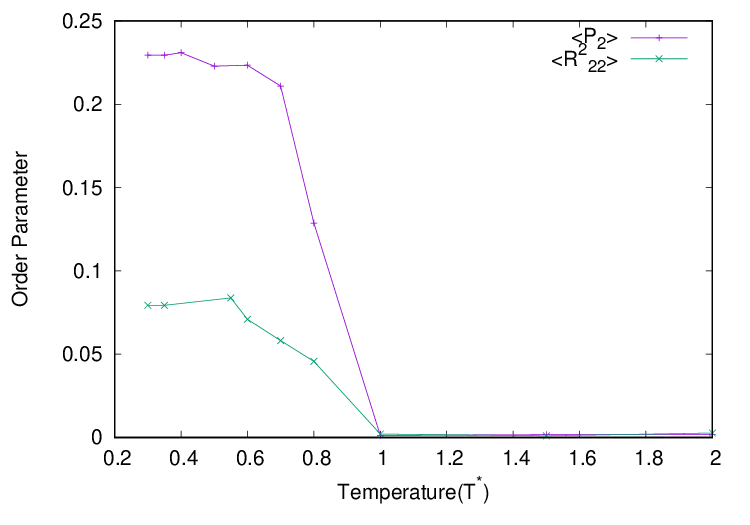}
        \caption{Twisted bilayered phase, $c=2.0$}
    \end{subfigure}
    %\hfill
    %\begin{subfigure}[b]{4cm}
    %    \includegraphics[width=\textwidth]{op_L35.5_c2.0.eps}
    %    \caption{Twisted cylinder, $c=2.0$}
    %\end{subfigure}
    %\hfill
    %\begin{subfigure}[b]{4cm}
    %    \includegraphics[width=\textwidth]{op_L42_c2.0.eps}
    %    \caption{Twisted vesicle, $c=2.0$}
    %\end{subfigure}
    \caption{Variation of the order parameters with temperature}
\end{figure*}
\\To quantify the degree of molecular ordering, orientational order parameters are calculated. These orientational order parameters are, $\langle P_{2} \rangle =(\frac{3}{2}\langle \cos^2\theta \rangle - \frac{1}{2}$) and $\langle R^{2}_{2,2} \rangle = \langle \frac{1}{2}(1+\cos^{2}{\theta})\cos2\phi \cos2\psi - \cos{\theta} \sin2\phi \sin2\psi \rangle$. Here, $\theta$, $\phi$, $\psi$ are the euler angles denoting molecular orientations in $3-d$ space. The plot of Fig.$5$ reveals, as the temperature($T^{*}$) is lowered, both the uniaxial order parameter $\langle P_{2} \rangle$ and biaxial order parameter $\langle R^{2}_{2,2} \rangle$ rise from zero (at high temperature) to positive non-zero values, indicating the onset of orientational ordering. The simultaneous growth of $\langle P_{2} \rangle$ and $\langle R^{2}_{2,2} \rangle$ implies, biaxial molecules trying to align their long-axes along layer normal while simultaneously organising their flat faces in a well ordered manner along layer plane. The plot in Fig.$5$ indicates, with increased molecular chirality average molecular alignment of the molecules along layer normal and along layer plane significantly reduces down due to the twist modification of the layered structures.%In case of the observed twisted phases, the phase biaxiality comes from the molecular flat face orientational ordering at low temperatures.With the variation of temperature, we have plotted the statistical averages of both the order parameters.  
%========================================================
\section*{Conclusions}
The importance for the simulation of hydrated complex chiral systems at mesoscopic scales cannot be overestimated. Due to the continued interest in chiral supramolecular assemblies{\cite{Changxia,Kataria}}, we have developed a solvent-free pair potential model to study various phase morphologies formed by chiral amphiphiles. Despite the coarse-grained nature and the absence of explicit water molecules, the emergence of hierarchical chiral order has been established with the increment of molecular inherent chirality. Our study captures the frustration arising from competition between molecular hydrophobicity and molecular intrinsic chirality, which drives instabilities in the observed morphological phases, disrupting the bilayer structure. The pitch of the twisted phases decreases with increasing chiral strength $c$, consistent with the continuum theory predictions. For a system of hydrated chiral amphiphiles, the fluidity decreases with increasing molecular chirality($c$) making the self-assembled system more viscous. The broken rotational symmetry about the long molecular axis provides a coherent orientational reference frame within which chiral torques accumulate constructively across neighboring molecules, enabling microscopic chirality to survive ensemble averaging and emerge as macroscopic twisted morphologies. The biaxial face ordering helps in the transfer of chirality from the molecular to the mesoscale. The emergence of twisted lamellae, cylindrical bilayers, and vesicular bilayers originates from the combined effects of chirality and molecular biaxiality. Beyond their relevance to synthetic materials, our findings carry implications for biological systems. Biological membranes are composed of homochiral phospholipids, and our results demonstrate that molecular-level homochirality alone is sufficient to drive spontaneous mesoscale twisted morphologies — providing a minimal physical basis for the prevalence of curved and helical membrane geometries in living cells. The molecular mechanisms underlying membrane rupture are an important topic for the investigations of programmed cell death{\cite{cell_deaths}}. We have shown that the fine competition between hydrophobicity and molecular intrinsic chirality can lead to the complete disruption of bilayer structures when chiral interaction dominates. Furthermore, the chirality-dependent reduction in molecular diffusivity observed across all three morphologies offers a mechanistic perspective on how chiral ordering of membrane lipids may regulate the mobility of membrane-embedded molecules, with implications for signal transduction and receptor dynamics. From a design standpoint, the existence of a tunable chirality threshold that controls bilayer stability and morphology suggests that synthetic chiral amphiphiles could be engineered to adopt specific morphologies — twisted vesicles, helical cylinders, or modulated lamellae — for targeted applications in drug delivery, where chirality-controlled permeability is desirable, and in tissue engineering scaffolds, where helical geometry enhances cell alignment and mechanical response. While the present coarse-grained model does not capture chemical specificity or electrostatic interactions present in real biological systems, it establishes the essential physical principle that molecular chirality is a sufficient and controllable parameter for programming mesoscale membrane architecture. Future work incorporating lipid shape asymmetry, mixed chirality systems, and charge interactions will further bridge the gap between these mesoscale simulations and the complexity of biological membrane environments.
%========================================================
\appendix
\renewcommand{\thesection}{\Alph{section}}

\section{Analytical Calculations of the Ellipsoid Contact Potential}
\label{app:ecp}
%\subsection*{Analytical calculations of ECP}
The total orientation-dependent energy strength function in our model system is given by,
\begin{equation}
%\begin{eqnarray}
\epsilon(\hat{a}_{i},\hat{a}_{j},\hat{r}) = \epsilon_0 \epsilon^{\nu}_1(\hat{a}_{i},\hat{a}_{j}) \epsilon^{\mu}_2(\hat{a}_{i},\hat{a}_{j},\hat{r})
%\end{eqnarray}
\end{equation}
Where, $\epsilon_0$ is a constant term and related to the energy scale of the system. The well-depth parameters $\mu$ \& $\nu$ can be varied for different compounds. Within total anisotropic energy function $\epsilon_1$ depends only on the orientations whereas $\epsilon_2$ depends on orientations of each ellipsoid as well as on the intermolecular separation between ellipsoidal centre of masses.
\\To calculate $\epsilon_2$ let us define a diagonal energy matrix ${\bf{\gamma}}$,
\begin{equation}
{\bf{\gamma}} = \begin{pmatrix}
(\frac{\epsilon_0}{\epsilon_x})^{\frac{1}{\mu}} & 0 & 0 \\
0 & (\frac{\epsilon_0}{\epsilon_y})^{\frac{1}{\mu}} & 0 \\
0 & 0 & (\frac{\epsilon_0}{\epsilon_z})^{\frac{1}{\mu}}
\end{pmatrix}
\end{equation}
Where, $\epsilon_x,\epsilon_y,\epsilon_z$ are respectively the well-depth values for side-side(s-s), width-width(w-w) and end-end(e-e) configurations of the neighbouring molecular ellipsoids. For this study, we have taken $\epsilon_x>\epsilon_y>\epsilon_z$ and $\epsilon_x=\epsilon_0$. Now, $\epsilon_2$ is given by,
\begin{equation}
\epsilon_{2}(\hat{a}_{i},\hat{a}_{j},\hat{r}) = \lambda_{E}(1-\lambda_{E})(\hat{r}^T.{\bf{H_E}}^{-1}.\hat{r}) 
\end{equation}
Here, $\lambda_E$ is a scalar parameter, which is evaluated by optimizing the above equation using Brent's method{\cite{Brent}}$\left(\lambda_E\in[0,1]\right)$. ${\bf{H_E}}$ is an affine combination of two matrices, ${\bf{A}}$ for i'th ellipsoid \& ${\bf{B}}$ for j'th ellipsoid, 
\begin{equation}
{\bf{H_E}} = \lambda_E{\bf{A}}+(1-\lambda_E){\bf{B}} 
\end{equation}
\\The matrices ${\bf{A}}$ \& ${\bf{B}}$ are given by,
\begin{equation}
{\bf{A}} = \hat{a}_{i}^{T}{\bf{\gamma}}\hat{a}_{i}\hspace{0.2cm};\hspace{0.2cm}{\bf{B}} = \hat{a}_{j}^{T}{\bf{\gamma}}\hat{a}_{j}
\end{equation}
\\Orientation dependent part of the total strength function $\epsilon_1(\hat{a}_{i},\hat{a}_{j})$ is given by{\cite{zanoni}},
\begin{equation}
\epsilon_{1}(\hat{a}_{i},\hat{a}_{j}) = \frac{1}{8}\sqrt{\frac{2\sigma_{y}}{\sigma_{x}}}\sigma^3_0\left[(\frac{\sigma_{y}}{\sigma_{x}})+(\frac{\sigma_{z}}{\sigma_{x}})^{2}\right]{\left|{\bf{H^{\prime}_E}}\right|^{-\frac{1}{2}}}
\end{equation}
In the above equation, ${\bf{H^{\prime}_E}}\hspace{-0.1cm}=\hspace{-0.1cm}({\bf{A}}+{\bf{B}})$. The values of the distances between two neighbouring molecular ellipsoids are $\sigma_x$, $\sigma_y$ and $\sigma_z$ when they are at s-s, w-w and e-e configurations respectively.
\\The distance of closest approach $\sigma(\hat{a}_{i},\hat{a}_{j},\hat{r})$ between two molecular ellipsoids is defined as,
\begin{equation}
%\begin{eqnarray}
\sigma^{-2}(\hat{a}_{i},\hat{a}_{j},\hat{r}) = \lambda_{D}(1-\lambda_{D})(\hat{r}^T.{\bf{H_{D}}}^{-1}.\hat{r}) 
%\end{eqnarray}
\end{equation}
The shape-related scalar parameter $\lambda_{D}$ is evaluated by optimizing the above equation using Brent's method{\cite{Brent}}, where the value of $\lambda_D\in[0,1]$. In the case of the distance function, ${\mathbf{H_{D}}}$ is also an affine combination of two matrices defined by two ellipsoids.
\begin{equation} 
{\mathbf{H_{D}}} = \lambda_{D}{\bf{M}} + (1-\lambda_{D}){\bf{N}}
\end{equation}
The form of the matrices are,
\begin{equation}
{\bf{M}} = \hat{a}_{i}^{T}{\bf{S}}^2\hat{a}_{i}\hspace{0.2cm};\hspace{0.2cm}{\bf{N}} = \hat{a}_{j}^{T}{\bf{S}}^2\hat{a}_{j}
\end{equation}
Where, ${\bf{S}}$ is a diagonal shape matrix.
\begin{equation}
{\bf{S}} = \begin{pmatrix}
\sigma_x & 0 & 0 \\
0 & \sigma_y & 0 \\
0 & 0 & \sigma_z
\end{pmatrix}
\end{equation}
$\sigma_x$, $\sigma_y$ \& $\sigma_z$ are molecular semi-axes lengths. In this study, we have taken systems for which $\sigma_x<\sigma_y<\sigma_z$ and $\sigma_x=\sigma_0$. Defining two vectors ${\bf{\kappa}}$ \& ${\bf{\kappa_{E}}}$ as ${\mathbf{H_{D}}}\cdot{\bf{\kappa}}={\mathbf{r}}$ \& ${\mathbf{H_{E}}}\cdot{\bf{\kappa_{E}}}={\mathbf{r}}$ and a box product term is denoted by ${\bf{b_p}}=\{(\hat{a}_{i3}\times\hat{a}_{j3})\cdot{\hat{r}_{ij}}\}(\hat{a}_{i3}\cdot\hat{a}_{j3})$. The expression of forces and torques are given below.
\\For $r\leq r_c$,
\begin{equation}
\begin{split}
\vec{f}_{ecp} & = \frac{8{\epsilon_0}{\epsilon^{\nu}_1}{\epsilon^{\mu}_2}}{\sigma_0r^{2}}[3r^{2}(2\rho^{-13}-\rho^{-7})\hat{r} \\
 & + 3{\sigma^{3}}{\lambda_D(1-\lambda_D)}(2\rho^{-13}-\rho^{-7})\{{\bf{\kappa}}-(\hat{r}.{\bf{\kappa}})\hat{r}\} \\ 
 & - {\lambda_E(1-\lambda_E)}{\mu}{\epsilon^{-1}_2}{\sigma_0}(\rho^{-12}-\rho^{-6}+\frac{1}{4})\{{\bf{\kappa_{E}}}-(\hat{r}.{\bf{\kappa_{E}}})\hat{r}\}] 
\end{split}
\end{equation}
\begin{equation}
\begin{split}
\vec{{\tau}}_{ecp} & = \frac{8{\epsilon_0}{\epsilon^{\nu}_1}{\epsilon^{\mu}_2}}{r^{2}}[-3{\sigma^{3}}{\lambda^{2}_D(1-\lambda_D)}\frac{(2\rho^{-13}-\rho^{-7})}{\sigma_0}\{{\bf{\kappa}}\cdot{\bf{M}}\times{\bf{\kappa}}\} \\ 
& - {\lambda^{2}_E(1-\lambda_E)}{\mu}{\epsilon^{-1}_2}(\rho^{-12}-\rho^{-6}+\frac{1}{4})\{{\bf{\kappa_{E}}}\cdot{\bf{A}}\times{\bf{\kappa_{E}}}\} \\
& + \frac{1}{2}\nu{r^{2}}(\rho^{-12}-\rho^{-6}+\frac{1}{4})\{\hat{a}_{i}^{T}\cdot({\bf{A}}+{\bf{B}})^{-1}\cdot\hat{a}_{i}\}] 
\end{split}
\end{equation}
For $r < r_c+w_f$,
\begin{equation}
\begin{split}
\vec{f}_{ecp} & = \frac{2{\epsilon_0}{\epsilon^{\nu}_1}{\epsilon^{\mu-1}_2}}{r^{2}}[{\lambda_E(1-\lambda_E)}{\mu}\{{\bf{\kappa_{E}}}-(\hat{r}.{\bf{\kappa_{E}}})\hat{r}\}]  
\end{split}
\end{equation}
\begin{equation}
\begin{split}
\vec{{\tau}}_{ecp} & = \frac{2{\epsilon_0}{\epsilon^{\nu}_1}{\epsilon^{\mu}_2}}{r^{2}}[-{\lambda^{2}_E(1-\lambda_E)}{\mu}{\epsilon^{-1}_2}\{{\bf{\kappa_{E}}}\cdot{\bf{A}}\times{\bf{\kappa_{E}}}\} \\
& + \frac{1}{2}\nu{r^{2}} \{\hat{a}_{i}^{T}\cdot({\bf{A}}+{\bf{B}})^{-1}\cdot\hat{a}_{i}\}] 
\end{split}
\end{equation}
For $r_c \leq r \leq w_f + w_c$,
\begin{equation}
\begin{split}
\vec{f}_{ecp} & = \frac{8{\epsilon_0}{\epsilon^{\nu}_1}{\epsilon^{\mu}_2}}{\sigma_0r^{2}}[3r^{2}(2\rho^{-13}_{_f}-\rho^{-7}_{_f})\hat{r} \\ 
 & \quad + 3{\sigma^{3}}{\lambda_D(1-\lambda_D)}(2\rho^{-13}_{_f}-\rho^{-7}_{_f})\{{\bf{\kappa}}-(\hat{r}.{\bf{\kappa}})\hat{r}\} \\ 
 & \quad - {\lambda_E(1-\lambda_E)}{\mu}{\epsilon^{-1}_2}{\sigma_0}(\rho^{-12}_{_f}-\rho^{-6}_{_f})\{{\bf{\kappa_{E}}}-(\hat{r}.{\bf{\kappa_{E}}})\hat{r}\}] 
\end{split}
\end{equation}
\begin{equation}
\begin{split}
\vec{{\tau}}_{ecp} & = \frac{8{\epsilon_0}{\epsilon^{\nu}_1}{\epsilon^{\mu}_2}}{r^{2}}[-3{\sigma^{3}}{\lambda^{2}_D(1-\lambda_D)}\frac{(2\rho^{-13}_{_f}-\rho^{-7}_{_f})}{\sigma_0}\{{\bf{\kappa}}\cdot{\bf{M}}\times{\bf{\kappa}}\} \\ 
& - {\lambda^{2}_E(1-\lambda_E)}{\mu}{\epsilon^{-1}_2}(\rho^{-12}_{_f}-\rho^{-6}_{_f})\{{\bf{\kappa_{E}}}\cdot{\bf{A}}\times{\bf{\kappa_{E}}}\} \\
& + \frac{1}{2}\nu{r^{2}}(\rho^{-12}_{_f}-\rho^{-6}_{_f}){\times} \{\hat{a}_{i}^{T}\cdot({\bf{A}}+{\bf{B}})^{-1}\cdot\hat{a}_{i}\}] 
\end{split}
\end{equation}
For the whole range of $0 \leq r \leq w_f + w_c$,
\begin{equation}
\begin{split}
\hspace{-0.5cm}\vec{f}_{chiral} & = \frac{4{c}{\epsilon_0}{\epsilon^{\nu}_1}{\epsilon^{\mu}_2}}{\sigma_0r^{2}}[-7r^{2}{\rho^{-8}}{\bf{b_p}}\hat{r} \\
 & - 7{\sigma^{3}}{\lambda_D(1-\lambda_D)}{\rho^{-8}}{\bf{b_p}}\{{\bf{\kappa}}-(\hat{r}.{\bf{\kappa}})\hat{r}\} \\ 
 & + 2{\lambda_E(1-\lambda_E)}{\mu}{\epsilon^{-1}_2}{\sigma_0}{\rho^{-7}}{\bf{b_p}}\{{\bf{\kappa_{E}}}-(\hat{r}.{\bf{\kappa_{E}}})\hat{r}\} \\ 
 & + {\rho^{-7}}{\sigma_0r^{2}}\{\frac{(\hat{a}_{i3}\times\hat{a}_{j3})\cdot(\hat{a}_{i3}\cdot\hat{a}_{j3})}{ \lvert \vec{r} \rvert }-\frac{\vec{r}}{r^{2}}{\bf{b_p}}\}] 
\end{split}
\end{equation}
%For $r\leq r_c$
\begin{equation}
\begin{split}
\hspace{-0.5cm}\vec{{\tau}}_{chiral} & = \frac{4{c}{\epsilon_0}{\epsilon^{\nu}_1}{\epsilon^{\mu}_2}}{\sigma_0r^{2}}[7{\sigma^{3}}{\lambda^{2}_D(1-\lambda_D)}{\rho^{-8}}{\bf{b_p}}\{{\bf{\kappa}}\cdot{\bf{M}}\times{\bf{\kappa}}\} \\ 
& - 2{\lambda^{2}_E(1-\lambda_E)}{\mu}{\epsilon^{-1}_2}{\sigma_0}{\rho^{-7}}{\bf{b_p}}\{{\bf{\kappa_{E}}}\cdot{\bf{A}}\times{\bf{\kappa_{E}}}\} \\ 
& - {\sigma_0r^{2}}{\nu}{\rho^{-7}}{\bf{b_p}}\{\hat{a}_{i}^{T}\cdot({\bf{A}}+{\bf{B}})^{-1}\cdot\hat{a}_{i}\} \\ 
& + {\rho^{-7}}{\sigma_0r^{2}}(\hat{a}_{i3}\times\{\frac{({\hat{a}_{j3}}\times{\vec{r}})(\hat{a}_{i3}\cdot\hat{a}_{j3})}{\vec{r}} \\
& + \{(\hat{a}_{i3}\times\hat{a}_{j3})\cdot{\hat{r}_{ij}}\}{\hat{a}_{j3}}\})] %{\frac{\partial {\bf{b_p}}}{\partial \hat{a}_{i3}}}*(-\hat{a}_{j3z}ryij + \hat{a}_{j3y}rzij)(\hat{a}_{i3}\cdot\hat{a}_{j3})
\end{split}
\end{equation}
%\end{widetext}
In the above expression of forces and torques, $\vec{f}_{ecp}$ \& $\vec{{\tau}}_{ecp}$ are van-der-waals force and torque mentioned in different regions. Whereas, $\vec{f}_{chiral}$ \& $\vec{{\tau}}_{chiral}$ are chiral force and torque.
% Paste the full cut content here — all text and equations
% from the former subsection
%========================================================
\section*{Data availability}
The data that support the findings of this study are available from the corresponding author upon reasonable request.

\section*{Code availability}
The simulation code used in this work is available upon reasonable request.

%========================================================
\section*{Acknowledgements}
S.M. acknowledges the support of Council of Scientific \& Industrial Research (CSIR), India, for providing Senior Research Fellowship. This work is supported by RUSA $2.0$ scheme of the University of Calcutta.

\section*{Author contributions}
S.M. performed the simulations. J.S. supervised the work. All authors discussed results and wrote the manuscript.

\section*{Competing interests}
The authors declare no competing interests.

%========================================================
\bibliographystyle{unsrt}

\begin{thebibliography}{99}
\bibitem{Castro} Vânia I. B. Castro, Rui L. Reis, Ricardo A. Pires, Iva. Pashkuleva. {\em Bioinspired helical systems with defined chirality assembled from discrete peptide and glycan amphiphiles}. Chem. Soc. Rev., {\bf 54}(18), 8325-8344, (2025).
\bibitem{Tschierske} C. Tschierske. {\em Mirror symmetry breaking in liquids and liquid crystals}. Liquid Crystals, {\bf 45}(13–15), 2221–2252, (2018).
\bibitem{Fellows} A.P. Fellows, B. John, M. Wolf, M. Thämer. {\em Spiral packing and chiral selectivity in model membranes probed by phase-resolved sum-frequency generation microscopy}. Nat Commun {\bf 15}, 3161 (2024).
\bibitem{Huang} Shuai Huang, Haifeng Yu, Quan Li. {\em Supramolecular Chirality Transfer toward Chiral Aggregation: Asymmetric Hierarchical Self-Assembly}. Advanced Science., {\bf 8}(8), (2021).
\bibitem{Wang} W. Wang, F. Chen, Z. Chen, Y. Sun, F. Liu and S. Zhang, {\em Unraveling hidden symmetry breaking in racemic compounds}, Proc. Natl. Acad. Sci. U.S.A. {\bf 123}(8), (2026).
\bibitem{Jiang} Hejin Jiang, Yuqian Jiang, Li Zhang, Zongxia Guo, Liu Minghua. {\em Symmetry Breaking and Amplification in a Self-Assembled Helix from Achiral trans-3-Nitrocinnamic Acid}. The Journal of Physical Chemistry C. {\bf 122}(23), 12559-12565, (2018).
\bibitem{Alaasar} M. Alaasar, AF. Darweesh, X. Cai, F. Liu, C. Tschierske. {\em Mirror Symmetry Breaking and Network Formation in Achiral Polycatenars with Thioether Tail}. Chemistry. {\bf 27}(60), 14921-14930, (2021).
\bibitem{Caimi} Federico Caimi, Giuliano Zanchetta. {\em Twisted Structures in Natural and Bioinspired Molecules: Self-Assembly and Propagation of Chirality Across Multiple Length Scales}. ACS Omega. {\bf 8}(20), (2023).
\bibitem{Ouyang} Guanghui Ouyang and Minghua Liu. {\em Self-assembly of chiral supra-amphiphiles}. Mater. Chem. Front. {\bf 4}(1), 155-167. (2020)
\bibitem{Schnur} Joel M. Schnur, {\em Lipid Tubules: A Paradigm for Molecularly Engineered Structures}. Science. {\bf 262}, 1669-1676. (1993).
\bibitem{Jonathan} Jonathan V. Selinger and Joel M. Schnur. {\em Theory of chiral lipid tubules}. Phys. Rev. Lett. {\bf 71}(24), 4091-4094. (1993).
\bibitem{Sawa} Y. Sawa, F. Ye, K. Urayama, T. Takigawa, V. Gimenez-Pinto, R.L.B. Selinger and J.V. Selinger, {\em Shape selection of twist-nematic-elastomer ribbons}, Proc. Natl. Acad. Sci. U.S.A. {\bf 108}(16). 6364-6368, (2011).
\bibitem{Barclay} Thomas G. Barclay, Kristina Constantopoulos, Wei Zhang, Michiya Fujiki, Nikolai Petrovsky and Janis G. Matisons. {\em Chiral Self-Assembly of Designed Amphiphiles: Influences on Aggregate Morphology}. Langmuir. {\bf 29}(32), 10001-10010. (2013).
\bibitem{Ziserman} Lior Ziserman, Amram Mor, Daniel Harries and Dganit Danino. {\em Curvature Instability in a Chiral Amphiphile Self-Assembly}. Phys. Rev. Lett. {\bf 106}(23), (2011)
\bibitem{Zeng} Wang Zeng, Yu Liu, Xianying Li, Wusong Jin and Dengqing Zhang. {\em Synthesis and self-assembly of chiral Gemini-shaped hexabenzocoronene amphiphiles}. Dyes and Pigments. {\bf 170}. (2019).
\bibitem{Wei} Y. Huang and Z. Wei. {\em Self-assembly of chiral amphiphiles with $\pi$-conjugated tectons}. Chin. Sci. Bull. {\bf 57}, 4246–4256. (2012).
\bibitem{Wu} Zheng Wu, Yun Yan and Jianbin Huang. {\em Advanced Molecular Self-Assemblies Facilitated by Simple Molecules}. Langmuir. {\bf 30}(48), 14375-14384. (2014).

\bibitem{Perram} John W. Perram and M.S. Wertheim, {\em Statistical mechanics of hard ellipsoids. I. Overlap algorithm and the contact function}, Journal of Computational Physics, {\bf 58}(3), 409-416, (1985).
\bibitem{Saupe} L. J. Yu and A. Saupe. {\em Observation of a Biaxial Nematic Phase in Potassium Laurate-1-Decanol-Water Mixtures}. Phys. Rev. Lett. {\bf 45}(12). (1980).
\bibitem{Saupe2} A. Saupe, P. Boonbrahm and L.J. Yu. {\em Biaxial nematic phases in amphiphilic systems}. J. Chim. Phys. {\bf 80}, 7-13. (1983).
\bibitem{Akpinar} E. Akpinar and A. M. Figueiredo Neto. {\em Experimental Conditions for the Stabilization of the Lyotropic Biaxial Nematic Mesophase}. Crystals, {\bf 9}(3), 158, (2019).
\bibitem{Reis} Dennys Reis, Erol Akpinar, Antônio Martins Figueiredo Neto. {\em Effect of Alkyl Chain Length of Alcohols on Cholesteric Uniaxial to Cholesteric Biaxial Phase Transitions in a Potassium Laurate/Alcohol/Potassium Sulfate/Water/Brucine Lyotropic Mixture: Evidence of a First-Order Phase Transition}. The Journal of Physical Chemistry B. {\bf 117}(3), 942-948. (2013).
\bibitem{Neto} A. M. Figueiredo Neto. {\em Micellar cholesteric lyotropic liquid crystals}. Liquid Crystals Reviews, {\bf 2}(1), 47–59. (2014).
\bibitem{Lubensky} Richard G. Priest, T. C. Lubensky. {\em Biaxial model of cholesteric liquid crystals}. Phys. Rev. A. {\bf 9}(2), 893-898. (1974).
\bibitem{Kroin} T. Kroin, A. M. Figueiredo Neto, L. Li\'ebert and Y. Galerne. {\em Chirality-induced biaxiality at the uniaxial-to-biaxial cholesteric phase transition}. Phys. Rev. A. {\bf 40}(8), 4647-4651. (1989).
\bibitem{Harris} A. B. Harris, R. D. Kamien, and T. C. Lubensky. {\em Microscopic origin of cholesteric pitch}. Phys. Rev. Lett. {\bf 78}(8), 1476-1479. (1997).
\bibitem{Brand} H.R. Brand and H. Pleiner. {\em Cholesteric to cholesteric phase transitions in liquid crystals}. J. Physique Lett. {\bf 46}(15), 711-718. (1985).

\bibitem{Shaoxuan} Shaoxuan Wang, Hanting Wang, Yunlong Zhao, Yuanyuan Wang and Lukang Ji. {\em Solvent-Driven Competition of Chirality Transfer in Cyanostilbene-Based Dipeptide Amphiphiles and Correlating Supramolecular Helicity with Cell Proliferation}. Langmuir. {\bf 42}(14), 9760-9770. (2026).
\bibitem{Rong-Ming} Rong-Ming Ho, Ming-Chia Li, Shih-Chieh Lin, Hsiao-Fang Wang, Yu-Der Lee, Hirokazu Hasegawa and Edwin L. Thomas. {\em Transfer of Chirality from Molecule to Phase in Self-Assembled Chiral Block Copolymers}. {\bf 134}(26), 10974-10986. (2012).

\bibitem{Feng} X. Wang and C. Feng. {\em Chiral fiber supramolecular hydrogels for tissue engineering}. WIREs Nanomedicine and Nanobiotechnology, {\bf 15}(2), (2023).
\bibitem{Tang} Zhen Du and Baoer Fan and Qiuju Dai and Lan Wang and Jia Guo and Zushan Ye and Naifu Cui and Jie Chen and Kun Tan and Ruixin Li and Wen Tang. {\em Supramolecular peptide nanostructures: Self-assembly and biomedical applications}. Giant. {\bf 9}. (2022)
\bibitem{Nanobiotechnol} Y. Wang, X. Zhang, K. Wan, N. Zhou, G. Wei and Z. Su. {\em Supramolecular peptide nano-assemblies for cancer diagnosis and therapy: from molecular design to material synthesis and function-specific applications}. J Nanobiotechnol. {\bf 19}(253). (2021). 
\bibitem{Paegel} J. Hu, WG. Cochrane, AX. Jones, DG. Blackmond, BM. Paegel. {\em Chiral lipid bilayers are enantioselectively permeable}. Nat Chem. {\bf 13}(8):786-791. (2021).
\bibitem{Sato} Kohei Sato, Wei Ji, Zaida Álvarez, Liam C. Palmer and Samuel I. Stupp. {\em Chiral Recognition of Lipid Bilayer Membranes by Supramolecular Assemblies of Peptide Amphiphiles}. ACS Biomater. Sci. Eng. {\bf 5}(6), 2786-2792. (2019).

\bibitem{Selinger} Robin L. B. Selinger, Jonathan V. Selinger and Anthony P. Malanoski and Joel M. Schnur. {\em Shape Selection in Chiral Self-Assembly}. Phys. Rev. Lett. {\bf 93}(15), (2004).
\bibitem{Selinger2} Jonathan V. Selinger and Joel M. Schnur. {\em Theory of chiral lipid tubules}. Phys. Rev. Lett. {\bf 71}(24), 4091-4094. (1993).
\bibitem{Seifert} Z. C. Tu and U. Seifert. {\em Concise theory of chiral lipid membranes}. Phys. Rev. E. {\bf 76}(3), (2007).
\bibitem{Nyrkova} I. A. Nyrkova and A. N. Semenov. {\em Twisted surfactant structures: an advanced theoretical model}. Soft Matter. {\bf 6}(3), 501-516. (2010).

\bibitem{Deserno} Ira R. Cooke, Kurt Kremer and Markus Deserno. {\em Tunable generic model for fluid bilayer membranes}. Phys. Rev. E. {\bf 72}(1). (2005).
\bibitem{Ugarte} D. Ugarte La Torre, S. Takada. {\em Coarse-grained implicit solvent lipid force field with a compatible resolution to the C$\alpha$ protein representation}. J Chem Phys. {\bf 153}(20). (2020).

%\bibitem{Nieh} Mu-Ping Nieh, V. A. Raghunathan, Charles J. Glinka, Thad A. Harroun, Georg Pabst and John Katsaras. {\em Magnetically Alignable Phase of Phospholipid “Bicelle” Mixtures Is a Chiral Nematic Made Up of Wormlike Micelles}. Langmuir. {\bf 20}(19). (2004).
\bibitem{Paramonov} L. Paramonov, SN. Yaliraki. {\em The directional contact distance of two ellipsoids: Coarse–grained potentials for anisotropic interactions}. J Chem Phys, {\bf 123}(19), (2005).
\bibitem{Perram_et_al} John. W. Perram, John Rasmussen, Eigil Pr\ae{}stgaard and Joel L. Lebowitz, {\em Ellipsoid contact potential: Theory and relation to overlap potentials}, Phys.Rev.E. {\bf 54}(6), (1996).
\bibitem{saha} J. Saha. {\em Soft ellipsoid potential for biaxial molecules}, Molecular Simulation, {\bf 42}(17), 1437-1443, (2016)
\bibitem{Bhowmick} S. Bhowmick, J. Saha. {\em Coarse-grained solvent-free model for computer simulation of self-assembled amphiphiles}. Sci Rep {\bf 15}, 40613 (2025).
\bibitem{Mondal} S. Mondal, J. Saha. {\em Computer simulation study of chiral liquid crystalline phases for biaxially shaped molecules}, Molecular Simulation, {\bf 52}(3), 179–189, (2026). 
\bibitem{Cooke} Ira R. Cooke, Markus. Deserno. {\em Solvent-free model for self-assembling fluid bilayer membranes: Stabilization of the fluid phase based on broad attractive tail potentials}, The Journal of Chemical Physics, {\bf 123}(22), 224710, (2005).
\bibitem{van der meer} B.W.van der Meer, G.Vertogen, A.J.Dekker, and J.G.J.Ypma , {\em A molecular-statistical theory of the temperature-dependent pitch in cholesteric liquid crystals}, The Journal of Chemical Physics. {\bf 65}, 3935, (1976).
\bibitem{paul} T. Paul, J. Saha. {\em Computer simulation study of novel chiral liquid crystal phases}, PHYSICAL REVIEW RESEARCH, {\bf 1}(3), (2019).
\bibitem{Brent} WH. Press, SA. Teukolsky, WT. Vellerling et al. {\em Numerical Recipes in Fortran}. 2nd ed. Cambridge: Cambridge University Press; 1992.
\bibitem{zanoni} R. Berardi and C. Fava and C. Zannoni, {\em A generalized Gay-Berne intermolecular potential for biaxial particles}, Chemical Physics Letters, {\bf 236}(4), 462-468, (1995).
\bibitem{Allen} MP. Allen, DJ. Tildesley. {\em Computer simulation of liquids}. Oxford: Clarendon Press ; 1987.
\bibitem{holian} Brad Lee Holian, Arthur F. Voter and Ramon Ravelo. {\em Thermostatted molecular dynamics: How to avoid the Toda demon hidden in Nos\'e-Hoover dynamics},Phys. Rev. E, {\bf 52}(3), 2338-2347, (1995).
\bibitem{Stone} A. J. Stone. {\em The description of bimolecular potentials, forces and torques: the S and V function expansions}. Molecular Physics, {\bf 36}(1), 241–256, (1978).
\bibitem{Frenkel} D. Frenkel, B. Smit. {\em Understanding Molecular Simulation}; Academic Press: New York, 1996.

\bibitem{Changxia} Changxia Liu, Dong Yang, Li Zhang and Minghua Liu. {\em Water inversed helicity of nanostructures from ionic self-assembly of a chiral gelator and an achiral component}. Soft Matter. {\bf 15}(32), 6557-6563. (2019). 
\bibitem{Kataria} M.Kataria, S.Seki. {\em Responsive Chirality: Tailoring Supramolecular Assemblies with External Stimuli as Future Platforms for Electronic/Spintronic Materials}. Chem. Eur. J. {\bf 31}(1). (2025). 
\bibitem{cell_deaths} Y. Zhang, X. Chen, C. Gueydan, J. Han. {\em Plasma membrane changes during programmed cell deaths}. Cell Research. {\bf 28}(1), 9-21, (2018).

\end{thebibliography}

\end{document}